\documentclass[fleqn,usenatbib]{mnras}
\usepackage{amsmath}	
\usepackage{newtxtext}
\usepackage{newtxmath}

\usepackage[T1]{fontenc}
\usepackage{ae,aecompl}

\usepackage{graphicx}	

\title[Torque formula for planetary migration]{Improved torque formula
for low and intermediate mass planetary migration}

\author[M. A. Jim\'enez \& F. S. Masset]{
Mar\'\i a Alejandra Jim\'enez \thanks{E-mail: mjimenez@icf.unam.mx}
and Fr\'ed\'eric S. Masset
\\
Instituto de Ciencias F\'\i sicas, Universidad Nacional Aut\'onoma de
M\'exico, Av. Universidad s/n, 62210 Cuernavaca, Mor., Mexico
}

\date{Accepted 2017 July 27. Received 2017 July 27; in original form 2017 April 25}

\pubyear{2017}

\begin{document}
\label{firstpage}
\pagerange{\pageref{firstpage}--\pageref{lastpage}}
\maketitle

\begin{abstract}
  The migration of planets on nearly circular, non-inclined orbits in
  protoplanetary discs is entirely described by the disc's
  torque. This torque is a complex function of the disc parameters,
  and essentially amounts to the sum of two components: the Lindblad
  torque and the corotation torque. Known torque formulae do not
  reproduce accurately the torque actually experienced in numerical
  simulations by low- and intermediate-mass planets in radiative
  discs. One of the main reasons for this inaccuracy is that these
  formulae have been worked out in two-dimensional analyses. Here we
  revisit the torque formula and update many of its dimensionless
  coefficients by means of tailored, three-dimensional numerical
  simulations. In particular, we derive the dependence of the Lindblad
  torque on the temperature gradient, the dependence of the corotation
  torque on the radial entropy gradient (and work out a suitable
  expression of this gradient in a three-dimensional disc). We also
  work out the dependence of the corotation torque on the radial
  temperature gradient, overlooked so far. Corotation torques are
  known to scale very steeply with the width of the horseshoe
  region. We extend the expression of this width to the domain of
  intermediate mass planets, so that our updated torque formula
  remains valid for planets up to typically several tens of Earth
  masses, provided these relatively massive planets do not
  significantly deplete their coorbital region. Our torque expression
  can be applied to low- and intermediate-mass planets in optically
  thick protoplanetary discs, as well as protomoons embedded in
  circumplanetary discs.
\end{abstract}


\section{Introduction} 
\label{sec:introduction}

An important ingredient for studies of planetary population synthesis
are the so-called migration maps, which are two-dimensional functions
$\Gamma(M_p,r)$ that provide the torque exerted by the disc on a
planet on a nearly circular orbit, as a function of its mass~$M_p$ and
of its distance~$r$ to the central star.  A migration map is intrinsic
to a given disc model. Over most of the $(M_p,r)$ domain, the torque
is generally a negative quantity, corresponding to an orbital decay of
the planet. Nevertheless, there may be some regions where the torque
is positive (named islands of outward migration). These regions
broadly occur where the disc's temperature drops faster than $r^{-1}$,
and span a mass range from a few Earth masses to potentially several
tens of Earth masses. The outcomes of models of planet population
synthesis depend sensitively on the outline of these islands
\citep[e.g.][]{2014A&A...569A..56C}. Migration maps can be established
in one of two ways: they can either be obtained by three dimensional
(3D) simulations, including all the physical ingredients needed for a
correct description of the disc \citep[e.g.][]{2015MNRAS.452.1717L},
or they can be obtained using torque formulae. The first approach is
naturally not suitable to the exploration of a large number of disc
models, owing to its considerable computational cost, and one must
resort to torque formulae \citep{2010ApJ...723.1393M,pbk11}. Numerical
simulations of low-mass planets embedded in radiative discs allows to
assess the validity of torque formulae. \citet{2010A&A...523A..30B}
have compared the torque experienced by a $20\;M_\oplus$ planet in a
radiative disc to the predictions of \citet{pbk11} and
\citet{2010ApJ...723.1393M}. The former displays a broad agreement
with the outcome of the simulations. The latter, however, was found to
be at odds with the simulations outcome, a discrepancy identified as
too saturated corotation torques, and later resolved when estimates of
the width of the horseshoe region in 3D discs became available
\citep{2016ApJ...817...19M}. Despite their broad agreement with
simulation outcomes, the torque formulae of \citet{pbk11} and
\citet{2010ApJ...723.1393M} lack the accuracy required by models of
planetary population synthesis. One of the main reasons is that these
analyses are based on results of two-dimensional analysis. Even though
they capture most of the mechanisms that contribute to the torque, the
dimensionless coefficient that they feature in many places may be
significantly off.

The purpose of this paper is to update the torque formula of
\citet{2010ApJ...723.1393M} by means of 3D numerical simulations,
successively tailored to determine the value of each of the different
dimensionless coefficients that we want to update. It is organised as
follows. In section~\ref{sec:focus-our-torque} we introduce and
discuss the different torque components that we want to update and we
describe the numerical code and setups that we use for this
purpose. In section~\ref{sec:results} we present our different
results, and provide an updated torque formula in
section~\ref{sec:new-torque-formula}. For the convenience of the
reader mainly interested in the formula, this section is
self-contained.  We compare the results of our updated torque formula
to published numerical simulations in section~\ref{sec:discussion},
and draw our conclusions in section~\ref{sec:conclusion}.

\section{Notation}
\label{sec:notation}
We consider a planet of mass $M_p$ on a circular orbit of radius $r_p$
around a central star of mass $M_\star$, with angular speed
$\Omega_p$, embedded in a disc of surface density $\Sigma(r)$ which
follows the power law:
\begin{equation}
  \label{eq:1}
  \Sigma(r)=\Sigma_0\left(\frac{r}{r_p}\right)^{-\alpha}.
\end{equation}
The disc's midplane temperature $T(r)$ obeys the law:
\begin{equation}
  \label{eq:2}
  T(r)=T_0\left(\frac{r}{r_p}\right)^{-\beta}.
\end{equation}
We denote with $\nu$ the kinematic viscosity of the disc, with $\chi$
its thermal diffusivity and with $\kappa$ its opacity.  The disc's
pressure scale length is $H$, and is given by:
\begin{equation}
  \label{eq:3}
  H=\frac{c_s}{\Omega_p},
\end{equation}
where $c_s$ is the disc's isothermal sound speed:
\begin{equation}
  \label{eq:4}
  c_s=\sqrt\frac{{\cal R}T}{\mu},
\end{equation}
${\cal R}$ being the constant of ideal gases and $\mu$ the mean
molecular weight. We will also make use of the disc's aspect ratio
\begin{equation}
  \label{eq:5}
  h=\frac Hr,
\end{equation}
and of the planet-to-star mass ratio
\begin{equation}
  \label{eq:6}
  q=\frac{M_p}{M_\star}.
\end{equation}

\section{Focus of our torque update}
\label{sec:focus-our-torque}
Both the Lindblad and corotation
torques can be normalised to:
\begin{equation}
  \label{eq:7}
  \Gamma_0=\Sigma_0\Omega_p^2r_p^4\left(\frac qh\right)^2=\Sigma_0\Omega_p^4r_p^6q^2c_s^{-2}.
\end{equation}
We draw hereafter a list of the different updates that we perform on
the normalised torques, which are dimensionless quantities.
\subsection{Lindblad torque}
\label{sec:lindblad-torque}
The normalised Lindblad torque depends on $\alpha$ and $\beta$, which
are respectively the slopes of surface density and temperature
\citep{ww86,1993Icar..102..150K}. Its dependence on $\alpha$ has
already been worked out by \citet{tanaka2002} by means of
semi-analytical calculations in the linear regime for planets in
globally isothermal discs, and we do not revise here this (weak)
dependence. On the contrary, its dependence on the temperature slope
has been worked out by \citet{2009arXiv0901.2265P} by a linear
analysis in two-dimensional discs with a softened potential.

Here we reevaluate the dependence of the Lindblad torque on the
gradient of midplane temperature. We undertake this analysis in
section~\ref{sec:depend-lindbl-torq}. We note that the Lindblad torque
can exhibit a dependence not only on $\alpha$ and $\beta$, but also on
$\alpha\beta$ \citep{1993Icar..102..150K}. This dependence is very
weak, however, and we neglect it. More generally, we seek, and
restrict ourselves to, linear dependencies of the different torque
components on the radial gradients of physical quantities at the
planet's location.

The Lindblad torque scales with the inverse squared of the sound speed
(see Eq.~\ref{eq:7}). It therefore depends on the thermal diffusivity
of the disc: when the latter is large enough that the disturbances
triggered by the planet at Lindblad resonances behave isothermally, it
scales with the inverse squared of the isothermal sound speed. On the
other hand, when the thermal diffusivity is small and the disturbances
behave adiabatically, the Lindblad torque scales with the inverse
squared of the adiabatic sound speed, and is therefore a factor of
$\gamma$ smaller, where $\gamma$ is the disc's adiabatic index. This
dependence has been considered by \citet{2010ApJ...723.1393M} and
\citet{pbk11}. We do not revise it here, and we will adopt for the
updated expression of the Lindblad torque (see
section~\ref{sec:general-case}) the dependence given by
\citet{2010ApJ...723.1393M}.

\subsection{Corotation torque}
\label{sec:corotation-torque-1}
\begin{figure*}
 \centering
 \includegraphics[width=.75\textwidth]{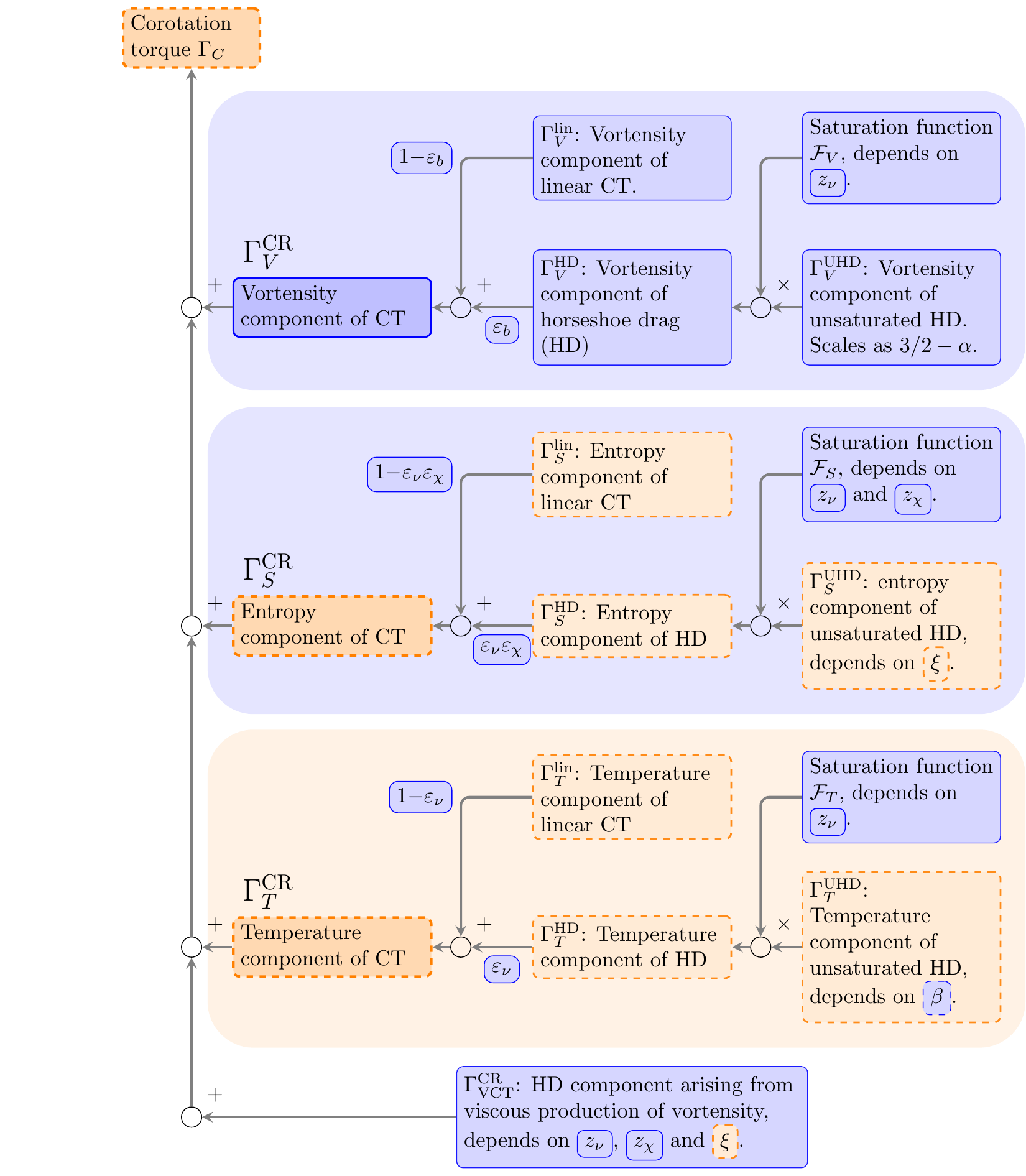}
 \caption{Synthetic representation of the different components of the
   corotation torque, showing how each of them is obtained (see
   section~\ref{sec:corotation-torque-1} for details). The blue
   rectangles show the components or variables which are not updated
   in this work. The orange rectangles (with a dashed frame to identify
   them on the printed version) show the variables or components for
   which we provide an updated version in this work. The component
   depending on the temperature gradient is integrated in a general
   torque formula for the first time, hence the orange background for
   this component. The weights $\varepsilon_b$, $1-\varepsilon_b$,
   etc. displayed next to the arrows of the linear torques and
   horseshoe drags are the blending coefficients of these
   components. Their expression, as the expression of $z_\nu$ and
   $z_\chi$, will be given in section~\ref{sec:new-torque-formula}. In
   the second frame (entropy torque), $\xi$ represents the new
   expression of the entropy gradient.}
 \label{fig:synth}
\end{figure*}

The corotation torque is the sum of three main components: one which
scales with the radial gradient of vortensity, one which scales with
the radial gradient of entropy, and one which scales with the radial
gradient of temperature. In addition, there is a contribution to this
torque arising from the vortensity viscously created at the contact
discontinuities which appear on the downstream separatrices in the
presence of an entropy gradient. These four contributions are
represented in Fig.~\ref{fig:synth}. Each of the first three
components is a blend of a linear component and of a horseshoe
drag. Each horseshoe drag is itself the product of the
\emph{unsaturated horseshoe drag} by a saturation function which is
generally a number between $0$ and $1$ (corresponding respectively to
a totally saturated and to an unsaturated horseshoe drag).  Each
torque component can in principle have its own saturation function.

We neither update, in this work, the saturation functions nor the
blending coefficients. \citet{2010ApJ...723.1393M} gave an expression
of the former, obtained from first principles by a two-dimensional
analysis of a simplified model of the horseshoe flow. The horseshoe
flow has since been shown to be essentially two-dimensional even in 3D
discs \citep{2016ApJ...817...19M}. On another hand, the flow arising
from turbulence in three dimensions and that arising from a kinematic
viscosity in a laminar 3D disc are markedly different
\citep{2011A&A...534A.107F}, so a 3D study of the torque saturation as
a function of kinematic viscosity would probably be pushing the
viscous model of protoplanetary discs beyond its domain of validity.

The blending coefficients were determined through a fit of numerical
experiments by \citet{2010ApJ...723.1393M} and by
\citet{pbk11}. \citet{2010ApJ...723.1393M} use the weights
$\varepsilon$ and $1-\varepsilon$ respectively for the horseshoe drag
and for the linear corotation torque (where $\varepsilon$ depends on
the torque component). \citet{pbk11} use the weights $1-K$ and $G$, so
that their sum is not necessarily $100$~\%. This allows them to
reproduce the fact that the horseshoe drag may be larger than its
unsaturated value. In the analysis of \citet{2010ApJ...723.1393M},
this fact is incorporated at a different place (namely the fact that
the saturation functions can have values slightly larger than
unity). Here we stick to the blending coefficients of
\citet{2010ApJ...723.1393M}.

The vortensity component of the corotation torque does not need any
update. \citet{2016ApJ...817...19M} have shown that the 3D horseshoe
drag in a barotropic disc has the same expression in a two dimensional
disc, and that the horseshoe region has a vertical separatrix, hence a
constant width on the whole vertical extent of the disc \citep[see
also][]{2015arXiv150503152F}. Naturally, an updated width of the
horseshoe region is used to evaluate this torque (more detail on this
is given in section~\ref{sec:width-hors-regi}), but no amendment has
to be done to the horseshoe drag expression. For the sake of
simplicity, we consider here that the gradient of
vortensity\footnote{More specifically, the quantity
  $d\log(\Sigma/B)/d\log r$, where $B$, the second Oort's constant, is
  half the flow's vertical vorticity.} is equal to $3/2-\alpha$. This
is true in any part of the disc where the surface density and angular
frequency are power laws of the distance to the star.

The temperature component of the corotation torque has been overlooked
in previous torque formulae. Here we obtain the value of this torque
as a by-product of the study of the Lindblad torque of
section~\ref{sec:depend-lindbl-torq} and present its properties in
section~\ref{sec:depend-corot-torq}. This torque component arises from
the production of vortensity by the radial temperature gradient, and
appears in locally isothermal calculations. The vortensity thus
produced is concentrated near the (downstream) separatrices. Although
it is not singular as the vortensity arising from non-barotropic
effects, which can formally be represented by a delta-function at the
separatrix \citep{mc09,pbck10}, it has nearly same spatial
distribution and it is reasonable to expect it to saturate in a
similar manner. We therefore assume that this torque has same
dependence on the viscosity as the entropy torque, and focus here
exclusively on the magnitude of the unsaturated horseshoe drag and
that of the linear torque.

Finally, we reevaluate the dependence of the entropy component on the
entropy gradient, a much needed determination since this component
plays a preponderant role in the appearance of outward migration
islands in migration maps. We also reevaluate how to determine the
entropy gradient in 3D disc. Whereas its expression is straightforward
in 2D discs, one may contemplate different ways of evaluating it in a
3D disc (such as the gradient in the midplane, or of the vertically
averaged quantity, etc.). Here we simply adopt the expression that
yields to a one-to-one relationship (with the least possible
dispersion) of the torque excess in a non-barotropic simulation with
respect to a barotropic simulation. This analysis is presented in
section~\ref{sec:torq-depend-radi}.

\subsection{Additional analysis}
\label{sec:additional-analysis}
In addition to the aforementioned reevaluations of key dependencies of
the torque, we have performed two side studies in order to improve the
accuracy and predictive power of the torque formula, which we present
below.

\subsubsection{Width of horseshoe region}
\label{sec:width-hors-regi-1}
The different components of the horseshoe drag scale with the width of
the horseshoe region to the fourth power. It is therefore crucial to
get an accurate estimate of this width. Besides, as soon as the ratio
$q/h^3$ is larger than $\sim 0.1$, the law that gives the width for
small mass planets
\citep{mak2006,2009arXiv0901.2263P,2016ApJ...817...19M} is no longer
valid and the horseshoe region is actually larger than what is
predicted by the low-mass law, resulting in a boost of the corotation
torque \citep{mak2006,2009arXiv0901.2263P,2015ApJ...806..182D}. The
threshold of $0.1$ is particularly stringent for typical
protoplanetary discs, and translates into $4\;M_\oplus$ only for a
disc with an aspect ratio of $h=0.05$ (which shows, incidentally, that
the comparisons mentioned in section~\ref{sec:introduction} were
performed much beyond the domain of validity of the torque formulae
for low-mass planets). In order to relax such a stringent condition on
the planetary mass, we study here the transition between the low-mass
regime, for which the width scales as $(q/h)^{1/2}$, and the high-mass
regime, for which, as in the restricted three body problem, the width
scales as $q^{1/3}$. This study is presented in
section~\ref{sec:width-hors-regi}.

\subsubsection{Thermal diffusivity}
\label{sec:thermal-diffusivity}
Since the saturation of the corotation torque depends on the value of
the disc's thermal diffusivity, it is important to determine this
value as accurately as possible. We have studied the radial spread of
an initially localised excess of temperature in order to assess the
accuracy of standard estimates which relate the disc's thermal
diffusivity to the disc's temperature, density and opacity. This study
is presented in section~\ref{sec:diff-coeff-}.

\subsection{Numerical details}
\label{sec:numerical-setups}

We conduct all our numerical experiments with the public hydrocode
FARGO3D\footnote{\url{http://fargo.in2p3.fr}}
\citep{2016ApJS..223...11B} with orbital advection enabled
\citep{fargo2000}. Here we use a spherical mesh that covers the full
azimuthal range $[-\pi,\pi]$. We denote $N_\phi$ the number of cells
in azimuth, $N_r$ the number of cells in radius and $N_\theta$ the
number of cells in colatitude. We always simulate a half-disc in
colatitude (since the planet is always coplanar with the disc), and
use reflecting boundary conditions at the midplane. All our
calculations involving a planet are performed in the corotating frame.

According to our needs, we either use a locally isothermal equation of
state, or we solve an energy equation. For the first case we use the
closure relationship between the pressure $p$ and density $\rho$:
\begin{equation}
  \label{eq:8}
  p(\mathbfit r,t)=c_s^2(r)\rho(\mathbfit r,t),
\end{equation}
where $c_s(r)$ is the isothermal sound speed, which depends on the
distance to the central star, and which is constant in time. In such
case there is no need to solve an energy equation. In the second case,
we solve the time evolution of the internal equation, and in the
absence of some form of thermal diffusion, the flow behaves
adiabatically. In this case we assume the gas to be ideal, so that the
pressure is given by:
\begin{equation}
  \label{eq:9}
  p(\mathbfit r, t)=(\gamma-1)e(\mathbfit r,t),
\end{equation}
where $e$ is the internal energy density.

Finally, in order to avoid the divergence of the planetary potential $\Phi_p$
in the vicinity of the planet, we soften it over a length scale
$\epsilon$:
\begin{equation}
  \label{eq:10}
  \Phi_p(\mathbfit{r})=-\frac{GM_p}{\sqrt{|\mathbfit{r}-\mathbfit{r}_p|^2+\epsilon^2}}.
\end{equation}
In three-dimensional calculations, $\epsilon$ has to be chosen small
compared to the pressure scale length, and larger or comparable to the
resolution. In all our numerical experiments we have $\epsilon=0.1H$.

\section{Results}
\label{sec:results}
We present hereafter the results of the numerical experiments
conducted to address the points listed in
section~\ref{sec:focus-our-torque}.

\subsection{Dependence of the Lindblad torque on the temperature
  gradient}
\label{sec:depend-lindbl-torq}
For this experiment, we use a locally isothermal setup, and use the
fact that at larger time the corotation torque saturates, so that the
total torque essentially amounts to the Lindblad torque.

In a locally isothermal setup, the Lindblad torque has
the form:
\begin{equation}
  \label{eq:11}
  \frac{\Gamma_{L}}{\Gamma_0}=-(2.34-0.099\alpha+k\beta),
\end{equation}
where as discussed in section~\ref{sec:lindblad-torque} the first two
terms of the right hand side come from the work of \citet{tanaka2002},
and where $k$ is the coefficient that we want to determine. In
principle, it suffices to perform two runs which have different values
of $\beta$ (all other parameters being the same) to infer $k$.

Our choice of parameters results from a trade-off between different
effects. We want the planetary mass to be sufficiently large so that
the horseshoe region is well resolved and can be saturated in a
reasonable amount of time. On the other hand, we want the planetary
mass to be sufficiently small so that the planet does not
significantly perturb the disc by carving a gap.

There is another difficulty inherent to this experiment: when the disc
is not globally isothermal ($\beta\ne 0$), it is prone to a baroclinic
instability \citep{2004ApJ...606.1070K} which results in considerable
noise in the torque curves, precluding any converged
measurement. However interesting this instability may be, it has for
us a parasitic character, and we aim at getting rid of it, while
preserving an almost complete saturation of the corotation torque. We
prevent the appearance of the instability by using some amount of
viscosity, and determine by dichotomy the minimal amount of viscosity
required to quench the instability. Also, the strength of the
instability scales with $|\beta|$, so one has to choose values of
$\beta$ sufficiently large to allow an accurate measurement of $k$,
but also sufficiently small to avoid the development of the baroclinic
instability, once the viscosity has been chosen.

With these considerations in mind, we have chosen for this numerical
experiment the following values: $q=2.4\cdot 10^{-5}$ (which
corresponds to $8\;M_\oplus$ if the central star has a solar mass) and
$\nu=7\cdot 10^{-8}r_p^2\Omega_p$. The disc's aspect ratio is
$h=0.05$. Although the kinematic viscosity is quite small, we do not
expect the planet to carve a significant gap in the disc over the time
scale required to achieve the saturation of the corotation torque. The
Hill radius of the planet [$r_p(q/3)^{1/3}$] is much smaller than the
pressure length scale, and the expression
$3h/[4(q/3)^{1/3}]+50\nu/(r_p^2\Omega_p q)$ is a factor of $2$ above
the critical value for gap opening \citep[see Eq.~(15)
of][]{crida06}. Even under these conditions a gap may eventually be
opened
\citep{2011ApJ...741...57D,2014ApJ...782...88F,2015ApJ...806L..15K},
but this will occur on time scales much longer than those of our
numerical experiments. As we shall see below, we do find evidence for
secular effects in our torque measurements, which affect our results
at the percent level.

  Our mesh extends from $\pi/2-3h$ to $\pi/2$ in colatitude, over
  $N_\theta=40$~cells, and from $0.6r_p$ to $1.4r_p$ in radius, over
  $N_r=200$~cells. In addition, we have $N_\phi=900$~cells. We use
  wave-killing boundary conditions in radius \citep{valborro06}, in
  order to avoid reflection of the wake at the radial boundaries. We
  therefore expect our results to be nearly insensitive to the exact
  location of the mesh radial boundaries, as these are located at a
  large number of pressure scale lengths from the orbit. We
  perform two simulations, with $\beta=1/2$ and $\beta=-1/2$
  respectively. Each one is run over $1000$~orbital periods of the
  planet. We choose $\alpha=3/2$ so that there is no vortensity
  gradient, and the only corotation torque present before saturation
  is the temperature component of the corotation torque.

\begin{figure}
  \centering
  \includegraphics[width=\columnwidth]{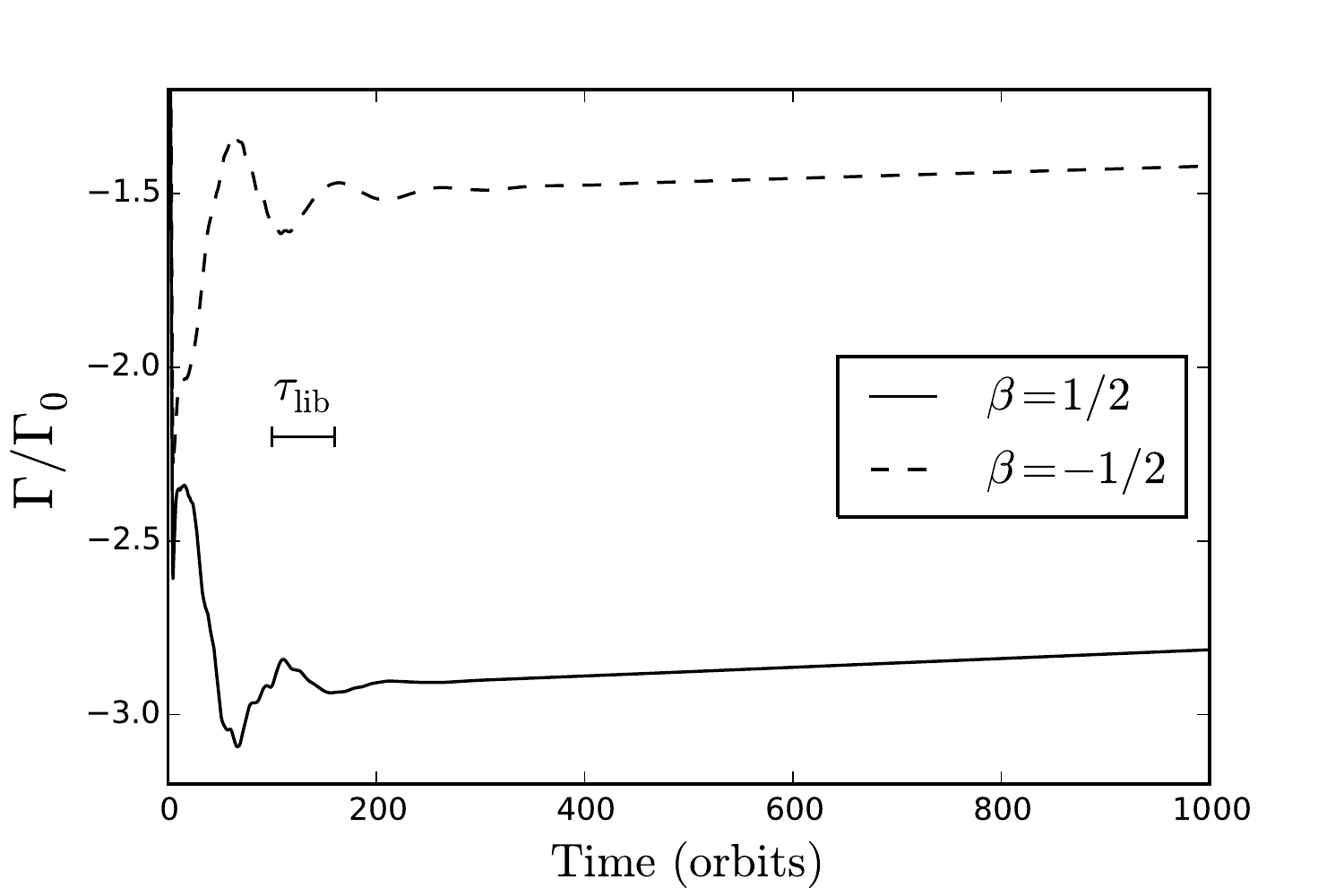}
  \caption{Total normalised torque on a planet with
    $q=2.4\cdot 10^{-5}$ for two different values of $\beta$, in a
    locally isothermal disc. The horizontal segment near $100$~orbits
    shows the horseshoe libration time. Past $\sim 300$~orbits, the
    corotation torque has saturated and the torque essentially amounts
    to the Lindblad torque. We see a systematic increase of the
    normalised torque of about $\sim 0.2$, in both cases, over
    $1000$~orbits.}
  \label{fig:lindgradt}
\end{figure}

This choice of parameters implies that the horseshoe region is
resolved over $11$~zones, which is sufficient to allow for a nearly
complete saturation of the torque \citep{mo2004}. We have
$q/h^3=0.19$, so that the flow is weakly non-linear. The horseshoe
libration time is $\sim 60$~orbits, so that our runs duration allows
for a full saturation of the corotation torque. The ratio of the
viscous time scale across the horseshoe region to the libration time
scale is $\sim 7$, which implies that the corotation should reach a
degree of saturation of $\sim 90$~\%. We show the torque as a function
of time for the two runs in Fig.~\ref{fig:lindgradt}. The corotation
torque saturates, leaving essentially the Lindblad torque. We see a
systematic, slow increase of the torque in both cases, which we
attribute to the slow opening of a very shallow dip (we find that the
rate of this increase decays as the viscosity increases).

\begin{figure}
  \centering
  \includegraphics[width=\columnwidth]{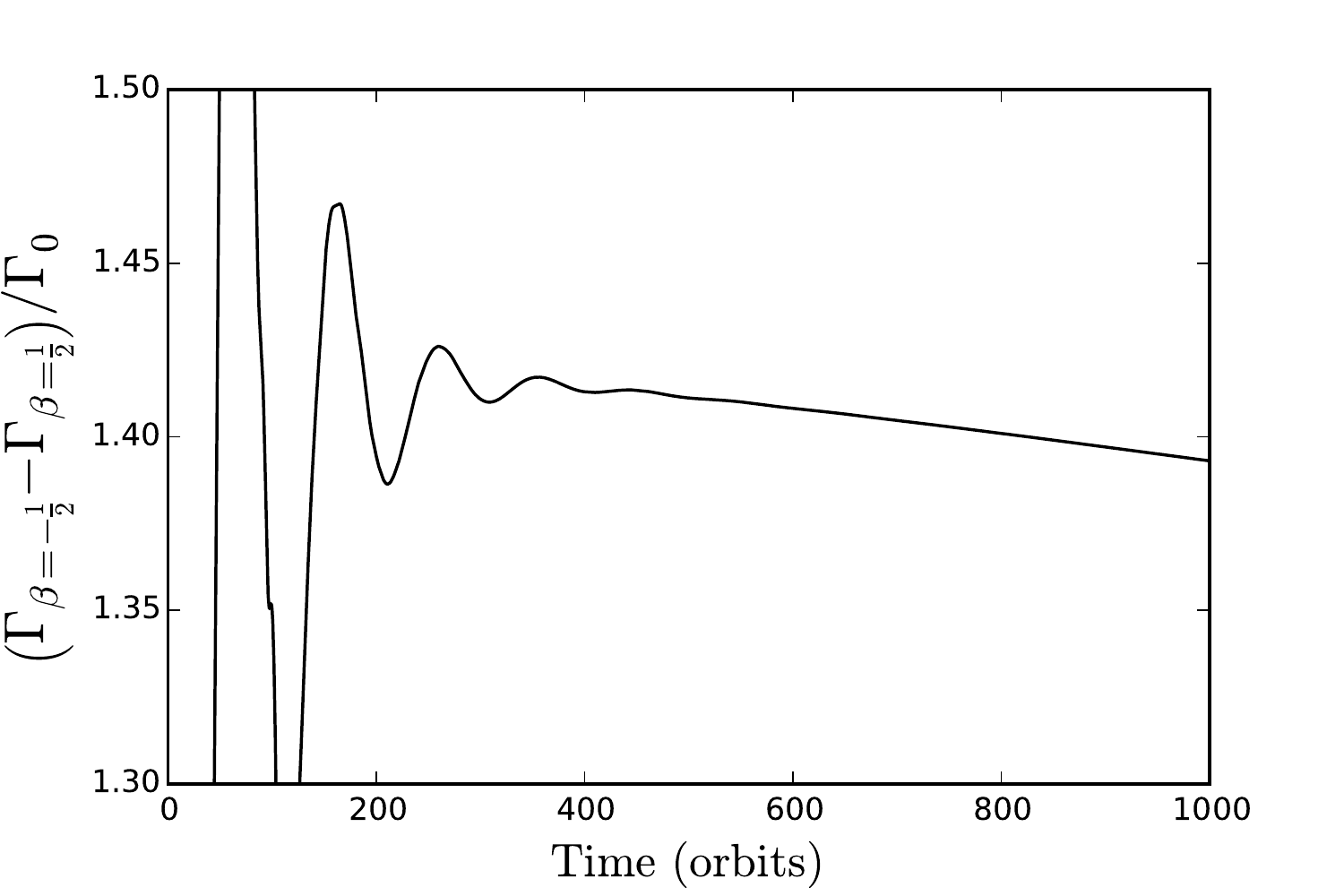}
  \caption{Result of the subtraction of the two curves of
    Fig.~\ref{fig:lindgradt}. The rising trends of the torques largely
    cancel out, and the difference shows an overall variation at the
    percent level.}
  \label{fig:klind}
\end{figure}

The value of $k$ can be directly obtained by subtracting the two
normalised torques. We show their difference in
Fig.~\ref{fig:klind}. From this figure we infer a value of
$k\approx 1.4$, comparable to, and marginally smaller than the value
of $1.7$ found by \citet{2009arXiv0901.2265P} for 2D discs.

\subsection{Dependence of the corotation torque on the temperature
  gradient}
\label{sec:depend-corot-torq}
We can further exploit the runs presented at the previous section to
work out the dependence on $\beta$ of $\Gamma _T^\mathrm{UHD}$ and
$\Gamma _T^\mathrm{lin}$ (see Fig.~\ref{fig:synth}).
Over the first half libration time, the torque is the sum of the
Lindblad torque and the unsaturated horseshoe drag. Since there is no
vortensity gradient in these runs, the horseshoe drag is entirely
attributable to the temperature gradient. As we now have an expression
of the Lindblad torque that takes the temperature gradient into
account, we can subtract this torque from the torque measured to get
an estimate of the corotation torque. We can also slightly improve
upon the results of the previous section by taking into account the
residual value of the corotation torque at larger time, as it does not
fully saturate.  We write the torque over the first half libration
time as:
\begin{equation}
  \label{eq:12}
  \Gamma^{(U)}_\beta=\Gamma_L+\Gamma _T^\mathrm{UHD},
\end{equation}
where the $(U)$ superscript stands for ``unsaturated'', and where the
$\beta$ subscript conveys that in this numerical experiment the only
parameter that is varied is $\beta$. In Eq.~\eqref{eq:12}, we do not
have a vortensity component for the reasons exposed above, and we do
not have an entropy component either because the setup is
isothermal. The planetary mass is sufficiently large, and the
viscosity sufficiently small, for the torque to be the (unsaturated)
horseshoe drag \citep{2009arXiv0901.2265P}.

\begin{figure}
  \centering
  \includegraphics[width=\columnwidth]{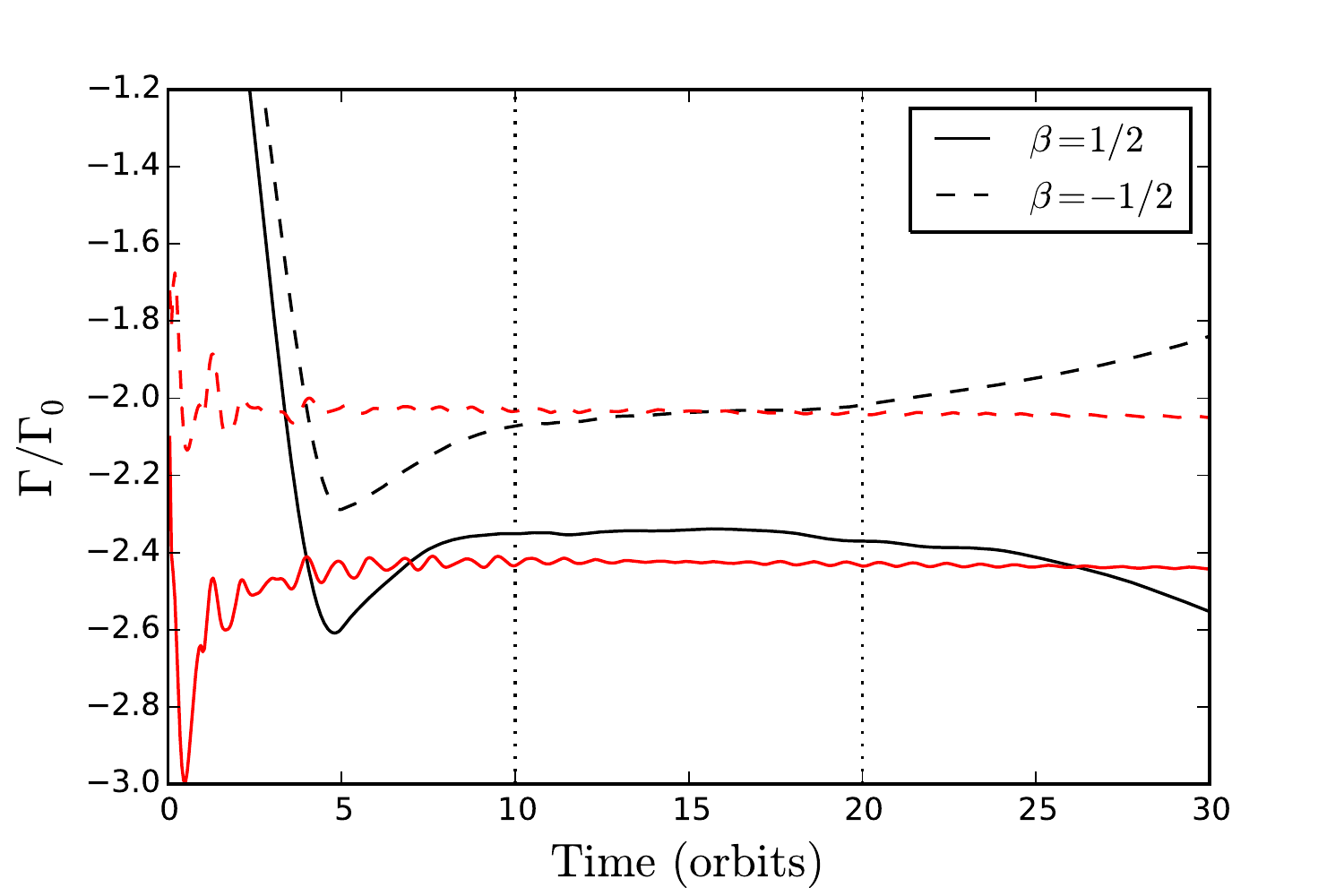}
  \caption{Torque as a function of time over the first half libration
    time, for the two runs presented in
    section~\ref{sec:depend-lindbl-torq} (black curves). The torques
    reach a plateau
    after $\sim 10$~orbits, and begin to depart from this plateau
    around $\sim 20$~orbits. The time averaged values between $10$~and
    $20$~orbits is $-2.34\Gamma_0$ for $\beta=1/2$ and $-2.04\Gamma_0$
    for $\beta=-1/2$. The grey curves (red in the electronic version)
    show the normalised torques for the same parameters, except that
    $q=3\cdot 10^{-6}$ and $\nu=10^{-5}r_p^2\Omega_p$, so that the
    corotation torque remains linear.}
  \label{fig:ctgradt}
\end{figure}

Writing the unsaturated corotation torque as
\begin{equation}
  \label{eq:13}
  \Gamma _T^\mathrm{UHD} =k'\beta\Gamma_0, 
\end{equation}
where $k'$ is the dimensionless coefficient that we
want to determine, we can write:
\begin{equation}
  \label{eq:14}
  \frac{\Gamma^{(U)}_\beta}{\Gamma_0}=K-k\beta+k'\beta,  
\end{equation} 
where $K=-2.34+0.099\alpha$ \citep{tanaka2002}. We can also write the
torque value at larger time as
\begin{equation}
  \label{eq:15}
  \frac{\Gamma^{\infty}_\beta}{\Gamma_0}=K-k\beta+\epsilon k'\beta, 
\end{equation}
where $\epsilon \ll 1$ represents the amount of saturation of the corotation
torque. From these relations we infer:
\begin{eqnarray}
  \label{eq:16}
  k-\epsilon k'&=&\frac{\Gamma^\infty_{-1/2}-\Gamma^\infty_{1/2}}{\Gamma_0}\equiv\frac{\Delta\Gamma^\infty}{\Gamma_0}\\
  \label{eq:17}
  k-k'&=&\frac{\Gamma^{(U)}_{-1/2}-\Gamma^{(U)}_{1/2}}{\Gamma_0}\equiv\frac{\Delta\Gamma^{(U)}}{\Gamma_0},
\end{eqnarray}
hence
\begin{eqnarray}
  \label{eq:18}
  k&=&\frac{\Delta\Gamma^\infty-\epsilon\Delta\Gamma^{(U)}}{\Gamma_0(1-\epsilon)}\\
  \label{eq:19}
  k'&=&\frac{\Delta\Gamma^\infty-\Delta\Gamma^{(U)}}{\Gamma_0(1-\epsilon)}.
\end{eqnarray}
We recover the fact that $k=\Delta\Gamma^\infty/\Gamma_0$ when
$\epsilon=0$, that is to say when the corotation torque is fully
saturated at larger time. Fig.~\ref{fig:ctgradt} shows the torque
behaviour at early time, from which we infer
$\Delta\Gamma^{(U)}=0.3\Gamma_0$. When $\epsilon=0$, this yields 
$k'=1.1$. Evaluating $k$ and $k'$ when $\epsilon=0.1$ (a reasonable
amount of residual corotation torque, as discussed in
section~\ref{sec:depend-lindbl-torq}), we get $k=1.5$, and
$k'=1.2$. This shows that the coefficients that we have obtained for
the temperature dependence of the normalised Lindblad and corotation
torques are accurate to within $\sim 0.1$.

We conclude this section by determining in a similar manner the linear
corotation torque. We simply have to amend the parameters so that the
corotation torque remains linear. This is achieved by adopting a much
smaller planet mass and a large viscosity
\citep{2009arXiv0901.2265P}. The results of these new runs are
displayed in Fig.~\ref{fig:ctgradt}, in which one can see that indeed
the torques remain nearly constant after a dynamical timescale and do
not show the variation toward a different value on $O(10)$~orbits
typical of the horseshoe drag \citep{2009arXiv0901.2265P}. These
results imply that the temperature component of the linear corotation
torque scales as:
\begin{equation}
  \label{eq:20}
  \Gamma_T^\mathrm{lin}=1.0\beta\Gamma_0,
\end{equation}
where again the numerical coefficient is determined to within
$\sim 0.1$.  At this stage, considering that we use the saturation
function and the weights provided by \citet{2010ApJ...723.1393M}, we
have completed our determination of the terms entering in
$\Gamma _T^\mathrm{CR}$ (see Fig.~\ref{fig:synth}, third frame).

\subsection{Torque dependence on the radial entropy gradient}
\label{sec:torq-depend-radi}
\begin{figure*}
  \centering 
  \includegraphics[width=.3\textwidth]{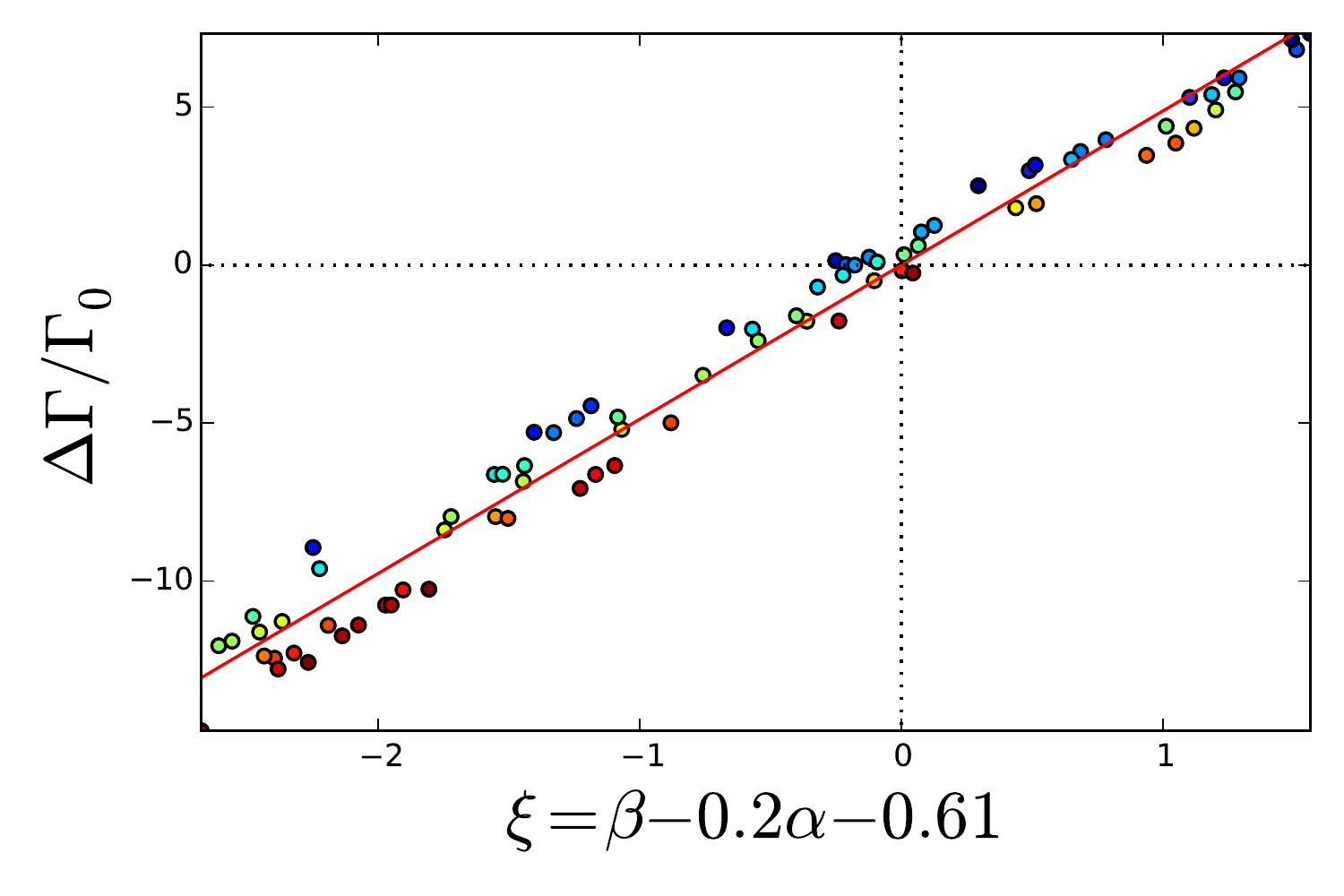}
  \includegraphics[width=.3\textwidth]{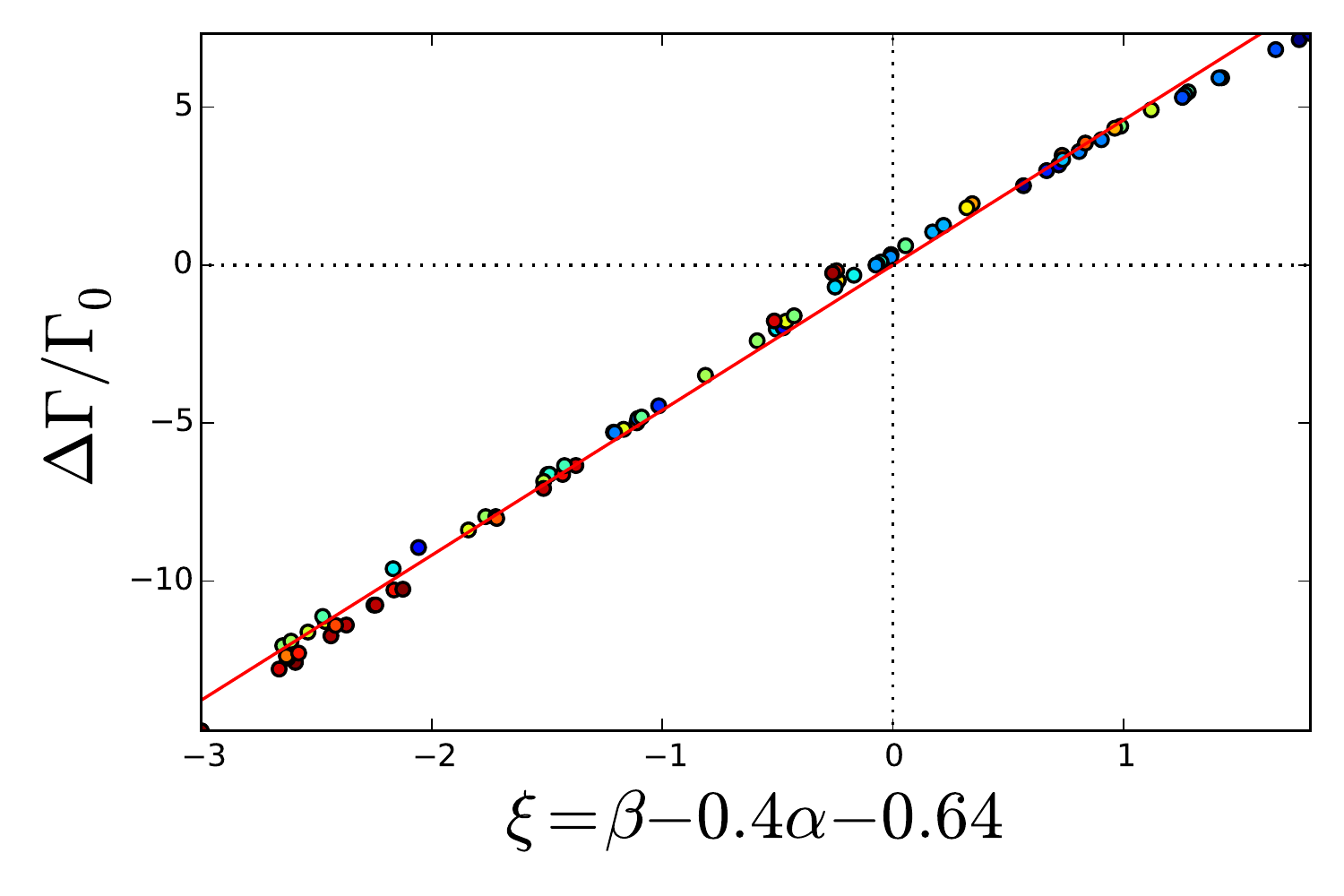}
  \includegraphics[width=.3\textwidth]{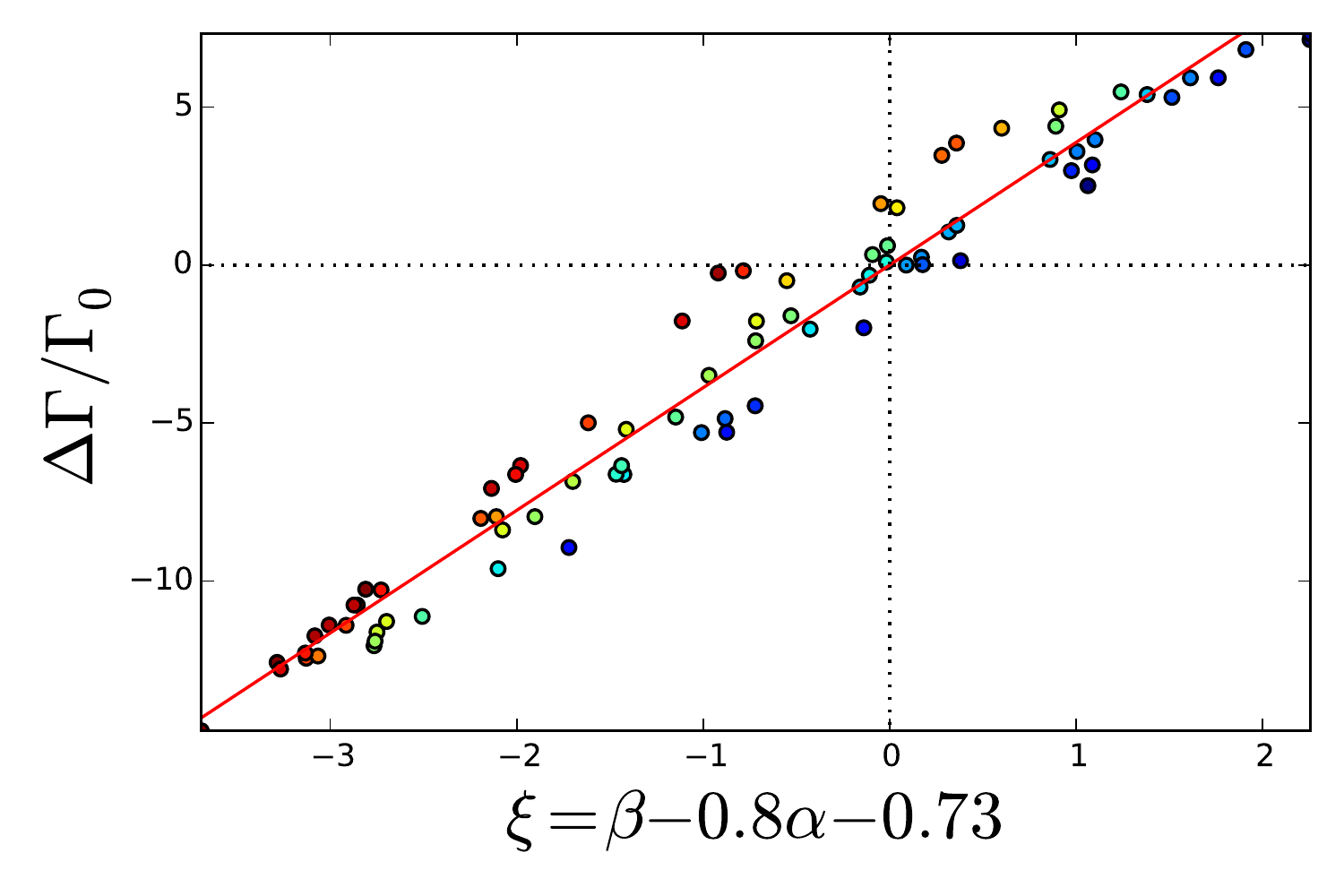}
  \caption{Torque difference $\Delta\Gamma$ between an adiabatic and
    isothermal setup, as a function of $\xi$, linear combination of
    $\beta$ and $\alpha$. We keep the coefficient of $\beta$ at the
    fixed value of one, and search the coefficient of $\alpha$ that
    minimises the sum of the squared distances to a linear fit. The
    first and third plots show cases for which a large scatter
    subsists, while the middle plot corresponds to the smallest sum of
    distances to the fit (solid line), hence to the best
    adjustment. In addition, we add a constant value to $\xi$ so that
    the linear fit of the data has no constant term, resulting in
    $\Delta\Gamma$ being proportional to $\xi$. In the electronic
    version, the points are coloured from blue to red according to the value of
    $\alpha$ (blue corresponds to $-1.5$ and red to $1.5$).}
  \label{fig:gradent3p}
\end{figure*}

We now perform a 3D version of the study presented by
\citet{2013LNP...861..201B} and update the terms entering in
$\Gamma_S^\mathrm{CR}$ (see Fig.~\ref{fig:synth},
second frame). We take an adiabatic index $\gamma=1.4$, unless
stated otherwise. We consider a large number of disc models (here $N=80$)
for which each value of the surface density and temperature slopes are
chosen randomly (namely, $\alpha$ is given by a random variable
uniformly distributed over $[-1.5,1.5]$, while $\beta$ is uniformly
distributed over $[-2,2]$). For each disc model, two runs are
performed: one with a locally isothermal equation of state, and one
with an adiabatic flow, which provide respectively the torque
$\Gamma_\mathrm{iso}$ and $\Gamma_\mathrm{adi}$. The parameters and
the duration of the runs are such that the corotation torque is in the
regime of unsaturated horseshoe drag. We then seek an appropriate
linear combination $\xi$ of $\alpha$ and $\beta$ such that the torque
excess, defined as
$\Delta\Gamma=\gamma\Gamma_\mathrm{adi}-\Gamma_\mathrm{iso}$, be a
one-to-one map of $\xi$. We illustrate this process in
Fig.~\ref{fig:gradent3p}. We find that an appropriate expression for
$\xi$ is the following:
\begin{equation}
  \label{eq:21}
  \xi=\beta-0.4\alpha-0.64,
\end{equation}
and that in this case we have:
\begin{equation}
  \label{eq:22}
  \Delta\Gamma=4.6\xi\Gamma_0.
\end{equation}

\emph{By construction,} our procedure implies that there is a linear
relationship between $\xi$ and the torque excess, and therefore the
calculation of the adiabatic torque is straightforward once one knows
the isothermal torque. In a 2D situation, the excess has been found to
scale with the radial entropy gradient \citep{bm08,pp08,mc09}. Here,
its expression resembles the gradient of entropy evaluated in the
midplane\footnote{Here this gradient is normalised so that the
  coefficient of $\beta$ is one.}, which is
$\beta-2(\gamma-1)/(3-\gamma)\alpha-3(\gamma-1)/(3-\gamma)=\beta-0.5\alpha-0.75$,
but differs slightly from it.  By extension, we call $\xi$ the radial
entropy gradient, but it must be kept in mind that this is an abuse of
language, as this quantity has not been obtained by considerations
about the entropy, but by a fit of numerical simulations. As
mentioned in section~\ref{sec:corotation-torque-1}, we assume that
this corotation torque component saturates as the entropy-related
torque in 2D discs.

We have not investigated in a systematic manner the
dependency of the different coefficients of Eq.~\eqref{eq:21} on the
adiabatic index. Nonetheless, for the reader interested in the torque
in a different kind of discs, we mentioned that we have conducted a
similar study for the case $\gamma=5/3$, and found in that
case that $\xi=\beta-0.58\alpha-0.85$ provides the one-to-one linear
relationship $\Delta\Gamma=4.1\xi\Gamma_0$.

At this stage we see that the horseshoe drag in the adiabatic case is
the sum of a term that scales with $\xi$ (combination of $\alpha$ and
$\beta$) and of the isothermal horseshoe drag, itself sum of a term
that scales with the vortensity gradient ($\frac 32-\alpha$ in power
law discs) and a term that scales with the temperature gradient
($\beta$) as found in section~\ref{sec:depend-corot-torq}. We have
three components and only two slopes ($\alpha$ and $\beta$). It may
seem desirable at first glance to simplify the expression for the net
corotation torque into a dependence on $\alpha$ and one on
$\beta$. This, as discussed by \citet{2013LNP...861..201B}, would
obfuscate the physical meaning of each term. Besides, each component
involves a different distribution of vortensity perturbation within
the horseshoe region, and thus in principle saturates in a different
manner, so the simplification could only be valid for the regime of
unsaturated horseshoe drag, not for the general regime. We shall
therefore keep the three components in the general expression of the
corotation torque.

We have performed a numerical experiment similar to the one described
above, with a lower mass ($q=10^{-6}$) and a higher viscosity
($\nu=2\cdot 10^{-5}r_p^2\Omega_p$), so that the corotation torque
remains linear \citep{2009arXiv0901.2265P}. We find in that case:
\begin{equation}
  \label{eq:23}
  \Delta\Gamma^\mathrm{lin}=0.8\xi\Gamma_0.
\end{equation}

\subsection{Width of the horseshoe region}
\label{sec:width-hors-regi}
As said in section~\ref{sec:width-hors-regi-1}, the different
components of the horseshoe drag scale with the width of the horseshoe
region to the fourth power. It is therefore important to have an
accurate value of this width, so we study its scaling as a function of
the planet mass in the regimes of low and intermediate masses,
corresponding respectively to $q\ll h^3$ and $q\lesssim h^3$ (we shall
provide slightly more quantitative definitions at the end of this
section). For this purpose we use globally isothermal simulations, and
perform a streamline analysis in the midplane, exploiting the fact
that the horseshoe region has a width which is independent of the
altitude in this case
\citep{2015arXiv150503152F,2016ApJ...817...19M}. Our runs have a setup
similar to the setup outlined in section~\ref{sec:depend-lindbl-torq},
except for the resolution $(N_\phi,N_r,N_\theta)=(1320,440,40)$. In
particular, we still have a null vortensity gradient. \citet{cm09}
have found that the horseshoe region is asymmetric when the
vortensity gradient is finite. Even though we average the width of the
horseshoe region in front and at the rear of the planet in order to
minimise the effect of an asymmetry, we prefer not to introduce a
source of asymmetry of the horseshoe region\footnote{The result of
  \citet{cm09} was obtained in two-dimensional discs. No similar
  result has been obtained in three-dimensional discs, so we assume
  that adopting a null vortensity gradient keeps the degree of
  asymmetry to an acceptable level.}. The planet mass is introduced
progressively over two orbital periods, and the width of the horseshoe
region is measured $20$~orbits after the introduction of the
planet. We perform a systematic study for three values of the aspect
ratio: $h=0.03$, $h=0.04$ and $h=0.05$. For each of these values, we
perform $30$ calculations with a planetary mass varying in a geometric
sequence from $0.05h^3$ to $4h^3$. The width of the horseshoe region
is determined automatically by dichotomy, averaging the results in
$\phi=1$~rad and $\phi=-1$~rad. We find that more accurate results are
obtained if the disc is inviscid when $q<0.4h^3$, and viscous for
larger masses. For lower mass indeed, a viscous drift of the disc can
distort severely the horseshoe region \citep{masset02}, whereas for
high masses, the planet can trigger vortices on the edges of its
horseshoe region if the disc is inviscid, which precludes an accurate
determination of the width of the horseshoe region.

\begin{figure}
  \centering
  \includegraphics[width=\columnwidth]{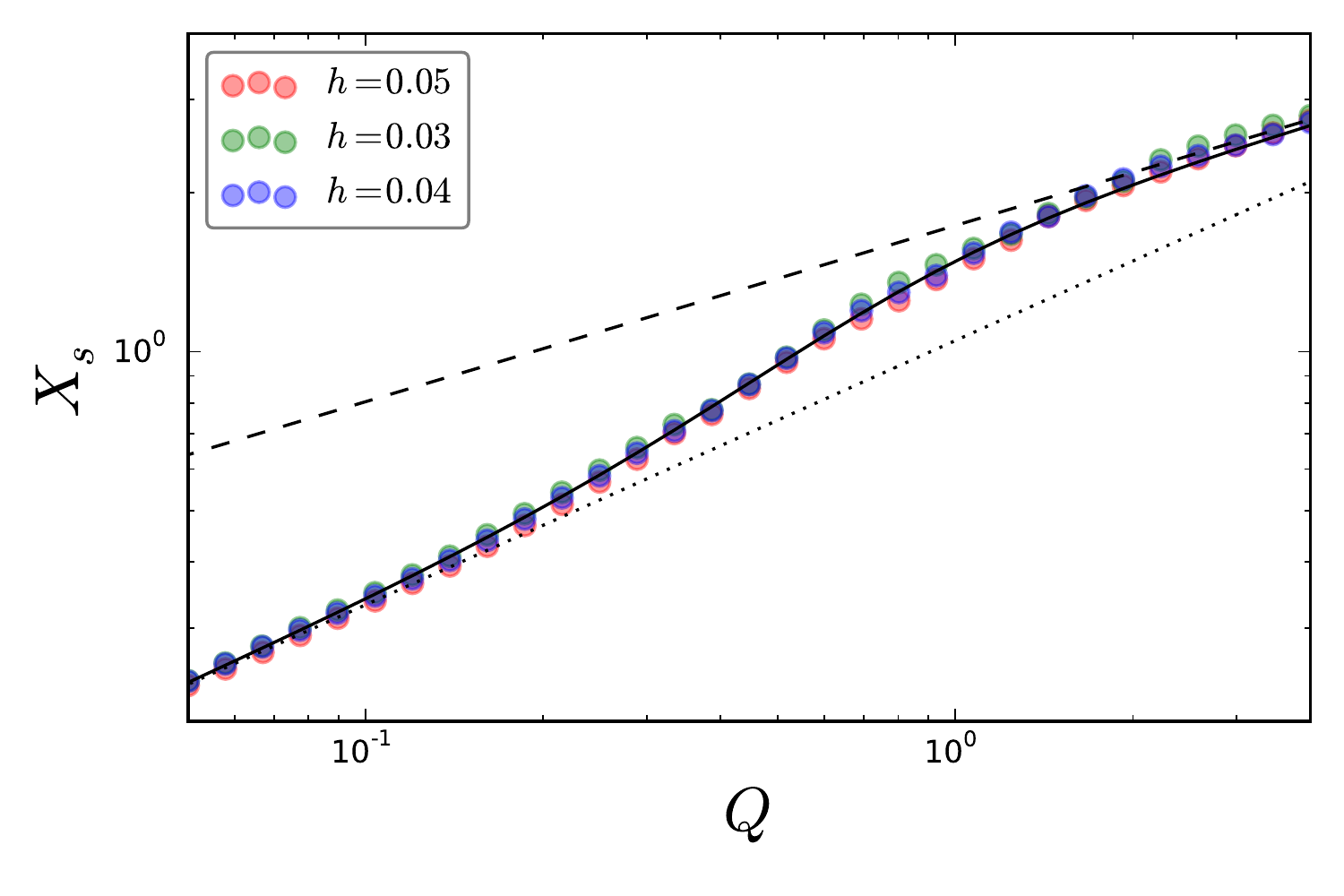}
  \caption{Normalised half width of the horseshoe region as a function
    of the mass in thermal masses. The data from the three different
    aspect ratios coincide nearly exactly. The solid curve corresponds
    to Eq.~\eqref{eq:29}, the dotted curve to Eq.~\eqref{eq:27} and
    the dashed curve to Eq.~\eqref{eq:28}. This figure should be
    compared to Fig.~7 of \citet{2015MNRAS.452.1717L}.}
  \label{fig:xsboostuni}
\end{figure}

For low masses, we have the scaling
\citep{2015MNRAS.452.1717L,2016ApJ...817...19M}:
\begin{equation}
  \label{eq:24}
  x_s=1.05r_p\sqrt\frac qh,
\end{equation}
whereas for large masses we recover the scaling found in 2D simulation
with softened potential \citep{mak2006}:
\begin{equation}
  \label{eq:25}
  x_s\approx 2.5r_p\left(q/3\right)^\frac13.
\end{equation}
Using the width normalised to the pressure length scale
$X_s=x_s/(hr_p)$ and denoting with $Q$ the planetary mass normalised
to the thermal mass $M_\mathrm{th}=c_s^3/(G\Omega_p)$:
\begin{equation}
  \label{eq:26}
  Q=\frac{M_p}{M_\mathrm{th}}=\frac{q}{h^3},
\end{equation}
we can rewrite Eqs.~\eqref{eq:24} and~\eqref{eq:25} as:
\begin{equation}
 \label{eq:27}
 X_s=1.05Q^{1/2}\mbox{~~~~when~}Q\ll 1
\end{equation}
and
\begin{equation}
 \label{eq:28}
 X_s=1.7Q^{1/3}\mbox{~~~~when~}Q\gg 1.
\end{equation}
The lack of an explicit reference to $h$ in these expressions suggests
that the normalised width of the horseshoe region has a universal
dependence on the mass expressed in thermal
masses. Fig.~\ref{fig:xsboostuni} confirms this expectation. We find
that a function that gives the asymptotic behaviours of
Eq.~\eqref{eq:27} and~\eqref{eq:28} and that matches satisfactorily
the width behaviour for intermediate masses is a blend of the
asymptotic values at low- and large-mass with weights respectively
$\epsilon$ and $1-\epsilon$, where $\epsilon=1/(1+2Q^2)$. This yields:
\begin{equation}
  \label{eq:29}
  X_s=\frac{1.05Q^{1/2}+3.4Q^{7/3}}{1+2Q^2},
\end{equation}
or
\begin{equation}
  \label{eq:30}
  x_s=\frac{1.05(q/h)^{1/2}+3.4q^{7/3}/h^6}{1+2q^2/h^6}r_p.
\end{equation}
From now on we will use Eq.~\eqref{eq:30} to evaluate the width of the
horseshoe region.

As a side result, this leads us to a reasonable definition of an
\emph{intermediate} mass protoplanet as a protoplanet that has a
horseshoe width intermediate between the widths given by the low-mass
and high-mass scaling (Eqs.~\eqref{eq:24}
and~\eqref{eq:25}). Examination of Fig.~\ref{fig:xsboostuni} shows
that this broadly corresponds to $0.2h^3\lesssim q\lesssim 2h^3$.

We assume that the corotation torque tends toward its linear value
when dissipation (either viscous or thermal) is large, regardless of
the planetary mass. This result has been shown by
\citet{2009arXiv0901.2265P} for low-mass planets, but it may be
questionable for intermediate mass objects. Although we have not
explored this issue in a systematic manner, we have performed
simulations with a planet of half the thermal mass ($Q=0.5$), and
found that it is subjected, at very large viscosity, to a vortensity
corotation torque comparable to the linear estimate. This result was
obtained considering globally isothermal discs with and without
vortensity gradients\footnote{The unsaturated {\em total} torque in
  isothermal simulations when $Q\sim 0.5$ can be off by several
  $\Gamma_0$ from its expected value, whereas we found with subsidiary
  calculations of isentropic discs that the total torque is correctly
  reproduced. These different behaviours have also been observed by
  \citet{2015arXiv150503152F,2017AJ....153..124F} and are linked to
  small scale features of the flow. Since the realistic discs in which
  we apply our torque formula are not isothermal, we believe that this
  discrepancy is unimportant, but it should be kept in mind when
  dealing with intermediate mass planets embedded in isothermal
  discs.}.  We note, however, that a substantial amount of dissipation
is required for a fall back of the corotation torque of an
intermediate mass object to its linear estimate. The decay toward the
linear regime is obtained for \citep[e.g.][]{2010ApJ...723.1393M}:
\begin{equation}
  \label{eq:31}
  \frac{hr_p\nu}{\Omega_px_s^3}\gg 0.1,
\end{equation}
and a similar condition holds, obtained by writing $\chi$ in stead of
$\nu$, for the decay of the entropy torque to its linear value. We
introduce the disc's alpha parameter $\alpha_\mathrm{ss}$
\citep{ss73}, and write
$\nu=\alpha_\mathrm{ss}r_p^2\Omega_ph^2$. Using for $x_s$ the
conservative value given by the low-mass estimate, we recast the
condition of Eq.~\eqref{eq:31} as:
\begin{equation}
  \label{eq:32}
\alpha_\mathrm{ss}  \gg 0.1Q^{3/2}.
\end{equation}
For $Q=0.2$, which corresponds to the mass at which departure from the
small mass regime becomes noticeable, this translates into
$\alpha_\mathrm{ss}\gg 10^{-2}$. This implies that under most
circumstances the corotation torque exerted on intermediate mass
objects should essentially be the horseshoe drag.

\subsection{Thermal diffusion coefficient}
\label{sec:diff-coeff-}
Previous estimates of the thermal diffusion coefficient have been
obtained by using values at the midplane of the disc
\citep[e.g.][]{pbk11, 2011A&A...536A..77B,2014A&A...564A.135B}. We
have undertaken simulations to check whether heat diffusion can indeed
be described by a diffusion coefficient evaluated at the disc's
midplane.

Following \citet{2009A&A...506..971K}, we write the thermal
diffusivity, in the absence of scattering, as (their Eq.~B.2):

\begin{equation}
  \label{eq:33}
  \chi=\frac{D}{c_v\rho}=\lambda\frac{4acT^3}{\rho^2c_v\kappa},
\end{equation}
where $D$ is the heat diffusion coefficient, $\lambda$, the
flux-limiter, is $1/3$ in the optically thick regions where the planet
is located, $a$ is the radiation constant and $c_v$ is the specific
heat at constant volume. We can recast Eq.~\eqref{eq:33} as:
\begin{equation}
  \label{eq:34}
  \chi=\frac{16(\gamma-1)\sigma T^3}{3\rho^2({\cal R}/\mu)\kappa},
\end{equation}
where $\sigma$ is Stefan's constant. We note that this expression has
oftentimes been transformed into an expression involving the disc's
thickness $H$ and orbital frequency $\Omega$ at the denominator. These
expressions, in general, overestimate by a factor $\gamma$ the thermal
diffusivity of Eq.~\eqref{eq:34}. The equation that describes the
hydrostatic equilibrium of the disc does not involve the adiabatic
index, and it yields the relationship $H=c_s/\Omega$, rather than
$H=c_s^\mathrm{adi}/\Omega=\sqrt\gamma c_s/\Omega$, independently of
the fact that an energy equation is used to describe the flow.
However, we suggest to simply use Eq.~\eqref{eq:34} rather than the
transformed expression, as the latter, which features all the
variables of Eq.~\eqref{eq:34} as well as new ones, does not bring any
simplification.

\begin{figure}
  \centering 
  \includegraphics[width=\columnwidth]{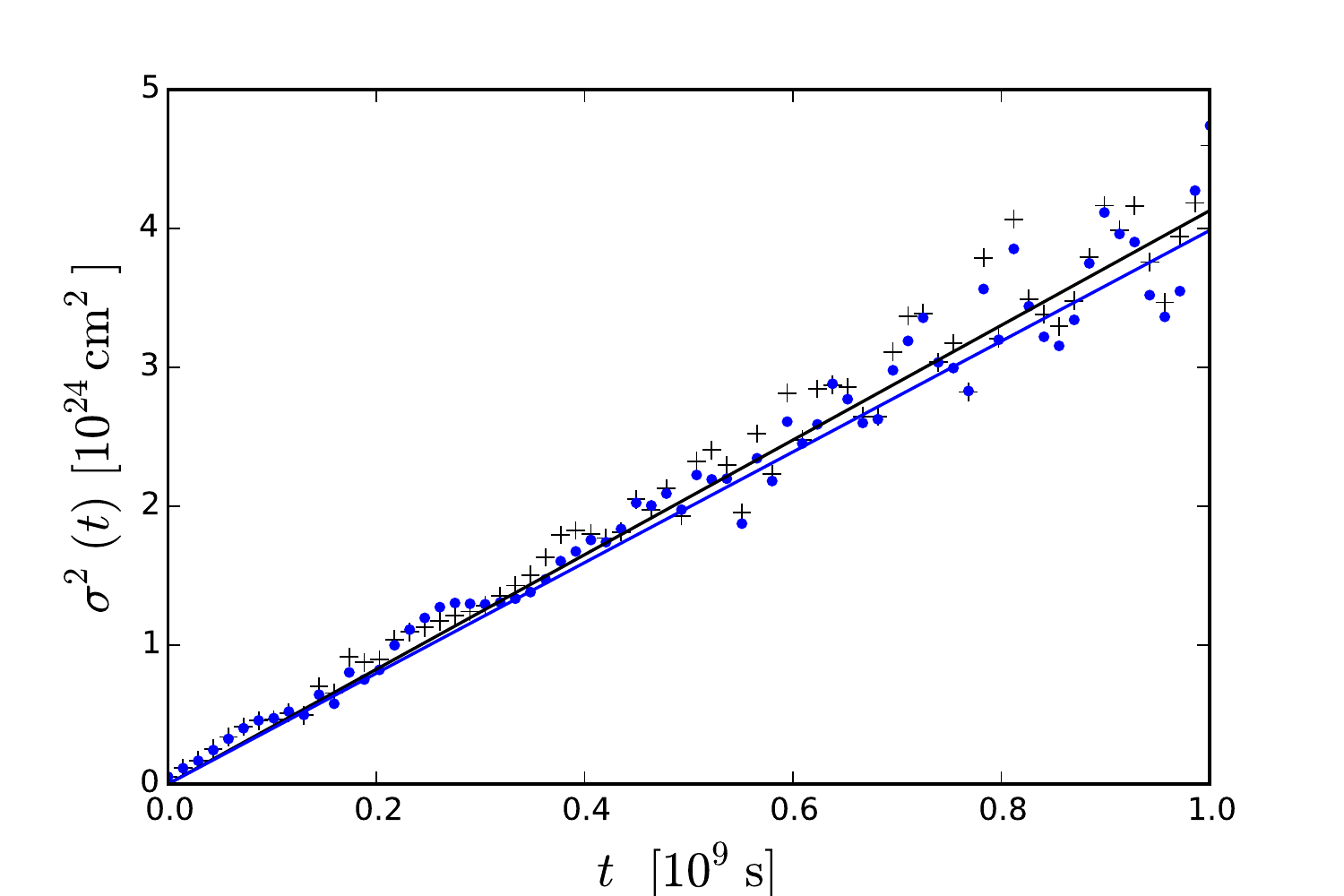}
  \caption{Mean square deviation of the temperature excess as a 
    function of time. The crosses are obtained by a measure of the 
    full width at half maximum (divided by $2\sqrt{\log 2}$) and the 
    dots (in blue in the electronic version) are obtained by means of 
    a second order polynomial fit of $\log\delta T$ on a 14-cell 
    interval centred on $r_0=1.5$~au. The slope obtained by a linear 
    regression fit of the crosses is $4.17\cdot 
    10^{15}$~cm$^2\,$s$^{-1}$, whereas the slope obtained by a linear 
    regression fit of the dots is $4.05\cdot 10^{15}$~cm$^2\,$s$^{-1}$.}
  \label{fig:tdiff}
\end{figure}

Here, we specifically study the \emph{radial} diffusion of heat, as it
is mainly this effect which determines (together with viscous
diffusion) the degree of saturation of the horseshoe drag.  Namely, we
study the radial spread of a radially localised temperature excess in
order to determine experimentally the thermal diffusivity of the disc
and compare it to the theoretical estimate.  In this section only we
use a module of radiative transfer in the flux limited diffusion
approximation \citep{1981ApJ...248..321L} with the flux limiter of
\citet{1989A&A...208...98K}, and solve the equation of radiative
energy in addition to that of internal energy (two-temperature
approach) as in \citet{2013A&A...549A.124B}.  We use a meridional mesh
of size $(N_\phi,N_r,N_\theta)=(1,800,43)$ spanning a radial range of
$0.75$~au to $3$~au, and a range in colatitude
$[\pi/2-0.1,\pi/2]$. The surface density at $1$~au is
$\Sigma_0=1700$~g$\,$cm$^{-2}$, the Rosseland opacity is fixed to
$1.8$~cm$^2\,$g$^{-1}$, and the kinematic viscosity is
$\nu=4.46\cdot 10^{14}$~cm$^2\,$s$^{-1}$.  The slope of surface
density is $\alpha=0.3$. The temperature profile of the disc at
equilibrium depends on a balance between viscous heating and radiative
losses through the disc's photospheres. We firstly run our initial
setup, which has an arbitrary aspect ratio, in order to relax the disc
toward hydrostatic and thermal equilibria. Once these are reached, we
increase by $1$~\% the temperature of a three-zone wide ring at
$r=1.5$~au (over the whole colatitude range), and restart the
simulation (hot ring run). We also perform a ``neutral'' restart in
which the temperature has not been altered. The radial profile of the
temperature excess obtained by subtracting the temperature profile of
the neutral run from the hot ring run (at the same date), quickly
adopts a nearly Gaussian profile:
\begin{equation}
  \label{eq:35}
  \delta T(r,t)\propto\frac{1}{\sigma(t)}\exp[-(r-r_0)^2/\sigma^2(t)].
\end{equation}
In linear diffusion theory, $\sigma^2(t)=4\chi t$. We use this
relationship to infer the value of $\chi$. Fig.~\ref{fig:tdiff} shows
the time behaviour of $\sigma^2$, which is nearly linear in time, as
expected. From the slope of the linear regression fit, we can estimate
that the value of $\chi$  at $r=r_0$ is:
\begin{equation}
  \label{eq:36}
  \chi\approx 1.03\cdot 10^{15}\;\mathrm{cm}^2\,\mathrm{s}^{-1}.
\end{equation}
An estimate based on the temperature excess averaged in colatitude,
rather than measured at the midplane, yields nearly exactly the same
result. At $r=1.5$~au, we have in the neutral run a midplane
temperature $T=509$~K, a density $\rho=4.33\cdot
10^{-10}$~g$\,$cm$^{-3}$, and ${\cal R/\mu}=3.615\cdot
10^7$~erg~K$^{-1}$~g$^{-1}$. Eqs.~\eqref{eq:34} yields a value $27$~\%
larger than the value found experimentally.

The disc of this numerical experiment is in radiative equilibrium and
its only source of heating is viscous friction. Its temperature
therefore decays from the midplane to low values near the boundaries
in colatitude. We have repeated this numerical experiment with a
nearly isothermal vertical profile, obtained by lowering the viscosity
by a factor of ten and by imposing a temperature at the upper boundary
$T_b$ such that $T_b^4=0.9T^4_\mathrm{midplane}$. The vertical heat
flux is therefore lowered by an order of magnitude with respect to the
previous disc, the temperature is nearly constant with respect to
colatitude, and the resulting disc has same midplane temperature as
the previous one. 

We measured a thermal diffusivity nearly equal to the one of the
previous disc, which shows that this quantity is rather insensitive to
the vertical profile of temperature, and determined by the midplane
temperature.

\section{New torque formula}
\label{sec:new-torque-formula}
We hereafter summarise our results and provide an updated torque
formula for different cases: the general case, the linear regime for
a radiative disc, and the linear regime for an isothermal disc.

\subsection{General case}
\label{sec:general-case}
Usually torque expressions are given in terms of the reference torque
$\Gamma_0$ of Eq.~\eqref{eq:7}. For the horseshoe drag, this
expression arises naturally when using the low mass value of the
horseshoe width (Eq.~\ref{eq:24}), and an horseshoe drag expression,
which involves the product $\Sigma_0\Omega_p^2x_s^4$. Here, however,
we use a more complex law (Eq.~\ref{eq:30}) for the horseshoe width,
so that casting our results in terms of $\Gamma_0$, for the different
components of the corotation torque, would lead to cumbersome
expressions. We therefore rather cast the different components of the
horseshoe drag in terms of $x_s$.

The total torque is a function of the slopes of surface density and
temperature $\alpha$ and $\beta$, of the disc's opacity, viscosity,
aspect ratio and surface density (respectively $\kappa$, $\nu$, $h$
and $\Sigma_0$) and of the planet orbital frequency, orbital radius
and mass to star mass ratio (respectively $\Omega_p$, $r_p$ and
$q$). It requires the prior evaluation of the thermal diffusivity
using Eq.~\eqref{eq:34}, and that of the half width of the horseshoe
region $x_s$, which is given, when the flow is adiabatic or behaves
nearly adiabatically over the horseshoe U-turn timescale, by a
modification of Eq.~\eqref{eq:30} that uses the adiabatic sound speed,
rather than the isothermal one:
\begin{equation}
  \label{eq:37}
  x_s=\frac{1.05(q/h')^{1/2}+3.4q^{7/3}/h'^6}{1+2q^2/h'^6}r_p,
\end{equation}
where $h'=h\sqrt{\gamma}$.

The total torque is the sum of the Lindblad torque $\Gamma_L$ and
corotation torque $\Gamma_C$:
\begin{equation}
  \label{eq:38}
  \Gamma_\mathrm{tot}=\Gamma_L+\Gamma_C.
\end{equation}
The Lindblad torque is given by:
\begin{equation}
  \label{eq:39}
  \Gamma_L=-(2.34-0.1\alpha+1.5\beta)\Gamma_0f(\chi/\chi_c),
\end{equation}
where the first two coefficients in the factor of the right hand side
($2.34$ and $-0.1$) are given by \citet{tanaka2002}, and where an
approximate value of the third coefficient (the factor of $\beta$) has
been obtained in section~\ref{sec:depend-lindbl-torq} and subsequently
slightly improved in section~\ref{sec:depend-corot-torq}. The function
$f(x)$ and critical diffusivity $\chi_c$ are given by
\citet{2010ApJ...723.1393M} and have respectively the expressions:
\begin{equation}
  \label{eq:40}
  f(x)=\frac{(x/2)^{1/2}+1/\gamma}{(x/2)^{1/2}+1},
\end{equation}
and:
\begin{equation}
  \label{eq:41}
  \chi_c=r_p^2h^2\Omega_p.
\end{equation}
The function $f(\chi/\chi_c)$ can be regarded as the inverse of an
effective adiabatic index $\gamma_\mathrm{eff}$ \citep[see
also][]{pbk11}.

We now turn to the evaluation of the four terms that compose the
corotation torque (see Fig.~\ref{fig:synth}). The vortensity component
$\Gamma_V^\mathrm{CR}$ of the corotation torque is given by:

\begin{equation}
  \label{eq:42}
  \Gamma_V^\mathrm{CR}=\varepsilon_b\Gamma_V^\mathrm{HD}+(1-\varepsilon_b)\Gamma_V^\mathrm{lin}.
\end{equation}
In this expression, the blending coefficient $\varepsilon_b$, as
advertised in section~\ref{sec:focus-our-torque}, has the expression
worked out by \citet{2010ApJ...723.1393M}:
\begin{equation}
  \label{eq:43}
  \varepsilon_b=(1+30hz_\nu)^{-1},
\end{equation}
where $z_\nu$ is given by \citet[][Eq.~79]{2010ApJ...723.1393M}:
\begin{equation}
  \label{eq:44}
  z_\nu=\frac{r_p\nu}{\Omega_px_s^3}.
\end{equation}
The (vortensity components of the) horseshoe drag
$\Gamma^\mathrm{HD}_V$ and linear corotation torque
$\Gamma_V^\mathrm{lin}$ are respectively:
\begin{equation}
  \label{eq:45}
  \Gamma_V^\mathrm{HD}={\cal F}_V\Gamma_V^\mathrm{UHD},
\end{equation}
and
\begin{equation}
  \label{eq:46}
  \Gamma_V^\mathrm{lin}=(0.976-0.640\alpha)\Sigma_0\Omega_p^2r_p^4(h')^{-2}=(0.976-0.640\alpha)\Gamma_0/\gamma,
\end{equation}
where the numerical coefficients of the right hand side of
Eq.~\eqref{eq:46} can be obtained from the data given by
\citet{tanaka2002}.  In Eq.~\eqref{eq:45} the two factors of the right
hand side are respectively the saturation function of the vortensity
component of the horseshoe drag, and the unsaturated horseshoe
drag. The former has the expression \citep{2010ApJ...723.1393M}:
\begin{equation}
  \label{eq:47}
  {\cal F}_V=\frac{8\pi}{3}z_\nu F(z_\nu),
\end{equation}
where $F(x)$ is defined as:
\begin{equation}
 \label{eq:48}
 F(x)= \left\{\begin{array}{ll}1-x^{1/2}&\mbox{~~~if $x<4/9$}\\
                4/(27x)& \mbox{~~~otherwise,}
\end{array}\right. 
\end{equation}
and the latter has the expression:
\begin{equation}
  \label{eq:49}
  \Gamma^\mathrm{UHD}_V=\frac 34\left(\frac 32-\alpha\right)\Sigma_0\Omega_p^2x_s^4.
\end{equation}
This last expression, initially established for two dimensional discs
\citep{wlpi91,masset01,cm09}, has been generalised to three
dimensional discs \citep{2016ApJ...817...19M}. Eqs.~\eqref{eq:42},
\eqref{eq:45}, \eqref{eq:49}, \eqref{eq:46} and~\eqref{eq:47}
correspond to the five cells of the first frame of
Fig.~\ref{fig:synth}, from left to right and bottom to top.

Next, we give the expression of the entropy component of the
corotation torque, corresponding to the second frame of
Fig.~\ref{fig:synth}. It reads\footnote{Note that in
  \citet{2010ApJ...723.1393M}, the thermal diffusivity was denoted by
  $\kappa$, whereas here it is denoted by $\chi$, while $\kappa$ is the
  opacity. As a consequence, $\varepsilon_\kappa$ ($z_\kappa$) has been
  changed into $\varepsilon_\chi$ ($z_\chi$).}:

\begin{equation}
  \label{eq:50}
  \Gamma_S^\mathrm{CR}=\varepsilon_\nu\varepsilon_\chi\Gamma_S^\mathrm{HD}+(1-\varepsilon_\nu\varepsilon_\chi)\Gamma_S^\mathrm{lin}.
\end{equation}
As for the vortensity torque, we keep for the blending coefficients
the expressions given by \citet{2010ApJ...723.1393M}:
\begin{equation}
  \label{eq:51}
  \varepsilon_\nu=[1+(6hz_\nu)^2]^{-1}
\end{equation}
and
\begin{equation}
  \label{eq:52}
  \varepsilon_\chi=(1+15hz_\chi)^{-1},
\end{equation}
where $z_\chi$ is defined in a similar fashion as $z_\nu$ \citep[see][Eq.~81]{2010ApJ...723.1393M}:
\begin{equation}
  \label{eq:53}
  z_\chi=\frac{r_p\chi}{\Omega_px_s^3}.
\end{equation}
The linear component of the torque, as seen in
section~\ref{sec:torq-depend-radi}, is:
\begin{equation}
  \label{eq:54}
  \Gamma_S^\mathrm{lin}=0.8\xi\Gamma_0/\gamma,
\end{equation}
where $\xi$ is given by:
\begin{equation}
  \label{eq:55}
  \xi=\beta-0.4\alpha-0.64.
\end{equation}
The horseshoe drag $\Gamma_S^\mathrm{HD}$ of Eq.~\eqref{eq:50} is
simply the product of the unsaturated horseshoe drag by the saturation
function of the entropy corotation torque:
\begin{equation}
  \label{eq:56}
  \Gamma_S^\mathrm{HD}={\cal F}_S\Gamma_S^\mathrm{UHD}.
\end{equation}
In two dimensions, the expression of the unsaturated horseshoe drag
can be written $K\xi\Sigma_0\Omega_p^2x_s^4$, where $K$ is a numerical
constant \citep{bm08}. We assume this dependency to hold in three
dimensions, and seek the value of $K$ using the results of
section~\ref{sec:torq-depend-radi}. In the numerical exploration that
we performed, we had, using Eq.~\eqref{eq:37},
$x_s=1.086r_p(q/h')^{1/2}$, from which we infer $K=3.3$. The
unsaturated torque expression is therefore:
\begin{equation}
  \label{eq:57}
  \Gamma_S^\mathrm{UHD}=3.3\xi\Sigma_0\Omega_p^2x_s^4.
\end{equation}
For the saturation function we keep the dependence given by
\citet{2010ApJ...723.1393M}:
\begin{equation}
  \label{eq:58}
  {\cal F}_S=1.2\times\overline{1.4z_\chi^{1/2}}\times\overline{1.8z_\nu^{1/2}},
\end{equation}
where $\overline{x}=\min(1,x)$. Eqs.~\eqref{eq:50}, \eqref{eq:56},
\eqref{eq:57}, \eqref{eq:54} and \eqref{eq:58} correspond to the five
cells of the second frame of Fig.~\ref{fig:synth}, from left to right
and bottom to top.

The third frame of Fig.~\ref{fig:synth} corresponds to the temperature
component of the corotation torque, that we now evaluate:
\begin{equation}
  \label{eq:59}
  \Gamma_T^\mathrm{CR}=\varepsilon_\nu\Gamma_T^\mathrm{HD}+(1-
\varepsilon_\nu)\Gamma_T^\mathrm{lin}.
\end{equation}
Since the vortensity distribution associated to this torque component
is roughly similar to that of the entropy torque (it is concentrated
near the separatrices), we assume that it behaves like the entropy
torque as a function of viscosity. However, we discard the dependence
on thermal diffusivity. Indeed, in the limit of a large diffusivity,
the disc behaves isothermally and the temperature torque subsists (as
we saw in section~\ref{sec:depend-corot-torq}), on the contrary of the
entropy torque. The linear component that features in
Eq.~\eqref{eq:59} reads (see section~\ref{sec:depend-corot-torq}):
\begin{equation}
  \label{eq:60}
  \Gamma_T^\mathrm{lin}=1.0\beta\Gamma_0/\gamma,
\end{equation}
while the horseshoe drag reads:
\begin{equation}
  \label{eq:61}
  \Gamma_T^\mathrm{HD}={\cal F}_T\Gamma_T^\mathrm{UHD}.
\end{equation}
From the above discussion we adopt:
\begin{equation}
  \label{eq:62}
  {\cal F}_T=1.2\times\overline{1.8z_\nu^{1/2}}.
\end{equation}
For the unsaturated horseshoe drag, we assume the form
$K'\beta\Sigma_0\Omega_p^2x_s^4$, i.e. similar to that of the entropy,
but with a different numerical constant and a scaling on the
temperature gradient. The toy model presented by \citet{cm09} suggests
it is reasonable, at least in the low mass limit, as it displays a
scaling in $x_s^2\Delta P$, where $\Delta P$ is the perturbation of
pressure at the stagnation point. This quantity, in the low mass
limit, scales as $\Sigma_0\Omega_p^2x_s^2$ if the stagnation point is
typically at a pressure scale length away from the planet. We seek the
value of $K'$ using the results of
section~\ref{sec:depend-corot-torq}. In that section, we had, using
Eq.~\eqref{eq:30} (rather than Eq.~\ref{eq:37} since the disc was
isothermal), $x_s=1.13r_p(q/h)^{1/2}$. Equating the form given above
with that of Eq.~\eqref{eq:13}, we infer $K'=k'/1.13^4=0.73$ and thus:
\begin{equation}
  \label{eq:63}
  \Gamma_T^\mathrm{UHD}=0.73\beta\Sigma_0\Omega_p^2x_s^4.
\end{equation}
Eqs.~\eqref{eq:59}, \eqref{eq:61}, \eqref{eq:63}, \eqref{eq:60} and
\eqref{eq:62} correspond to the five cells of the third frame of
Fig.~\ref{fig:synth}, from left to right and bottom to top.

In a fourth and last stage, we evaluate the additional component,
arising from the viscous creation of vortensity at the abrupt density
jumps that appear at the contact discontinuities at the separatrices
of the horseshoe region. For brevity and definiteness we hereafter
call this term the viscous coupling term (VCT).
\citet{2010ApJ...723.1393M} give an expression of this term (the
second part of the bracket of their Eq.~129). This term corresponds to
an asymptotic value (at larger time) and therefore already embeds its
own saturation function. It naturally scales with the entropy
gradient, as it is this gradient which primarily determines the
magnitude of the density jumps at the separatrices. As emphasised in
section~\ref{sec:torq-depend-radi}, the quantity $\xi$ was determined
from a best fit of numerical data, and named entropy gradient on the
grounds of the very close resemblance of the effect studied in that
section with the well-studied two-dimensional process that triggers
the appearance of an additional torque component that scales with the
radial entropy gradient \citep{pp08,bm08}. We therefore suggest that
the three-dimensional version of the VCT scales with $\xi$.  Our
choice of normalisation of $\xi$ is that the coefficient of $\beta$ is
unity, whereas it is $-1/\gamma$ with the normalisation of
\citet{2010ApJ...723.1393M}. As this torque behaves as a bulk term (it
implies a smooth distribution of vortensity from corotation to the
separatrices), and since it has no known linear equivalent, we follow
\citet{2010ApJ...723.1393M} and assume it to decay like the vortensity
component of the horseshoe drag, proportionally to $\varepsilon_b$:
\begin{equation}
  \label{eq:64}
  \Gamma_\mathrm{VCT}^\mathrm{CR}=\frac{4\pi\xi}{\gamma}\Sigma_0\Omega_p^2x_s^4\epsilon_bz_\nu\left[\frac{z_\nu
      F(z_\nu)-z_\chi F(z_\chi)}{z_\nu-z_\chi}\right]
\end{equation}
The full corotation torque is finally given by:
\begin{equation}
  \label{eq:65}
  \Gamma_C=\Gamma_V^\mathrm{CR}+\Gamma_S^\mathrm{CR}+\Gamma_T^\mathrm{CR}+\Gamma_\mathrm{VCT}^\mathrm{CR}.
\end{equation}
Eqs.~\eqref{eq:37} to~\eqref{eq:65}, together with Eq.~\eqref{eq:34},
provide an expression for the total tidal torque exerted on a low to
intermediate mass planet in circular orbit in an optically thick disc in hydrostatic
and radiative equilibrium.

\subsection{Linear regime in the general case}
\label{sec:linear-regime}
We give hereafter the simpler expression of the torque in the linear
regime. It reduces to:
\begin{equation}
  \label{eq:66}
  \Gamma_\mathrm{tot}^\mathrm{lin}=\Gamma_L+\Gamma_V^\mathrm{lin}+\Gamma_S^\mathrm{lin}+\Gamma_T^\mathrm{lin},
\end{equation}
which simplifies as:
\begin{equation}
  \label{eq:67}
  \Gamma_\mathrm{tot}^\mathrm{lin}=\Gamma_L+(0.46-0.96\alpha+1.8\beta)\frac{\Gamma_0}\gamma,
\end{equation}
where $\Gamma_L$ is given by Eq.~\eqref{eq:39}.

\subsection{Linear regime in an isothermal disc}
\label{sec:linear-regime-an}
We can further simplify this expression in the case of an isothermal
disc (in which case we discard the entropy component of the
corotation torque). This yields:
\begin{equation}
  \label{eq:68}
  \Gamma_\mathrm{tot}^\mathrm{lin,iso}=-(1.36+0.54\alpha+0.5\beta)\Gamma_0.
\end{equation}
This result is in good agreement with the result obtained by
\citet[][see their Eq. 9]{2010ApJ...724..730D} in three-dimensional
simulations of locally isothermal discs.

\begin{figure*}
  \includegraphics[width=.32\textwidth]{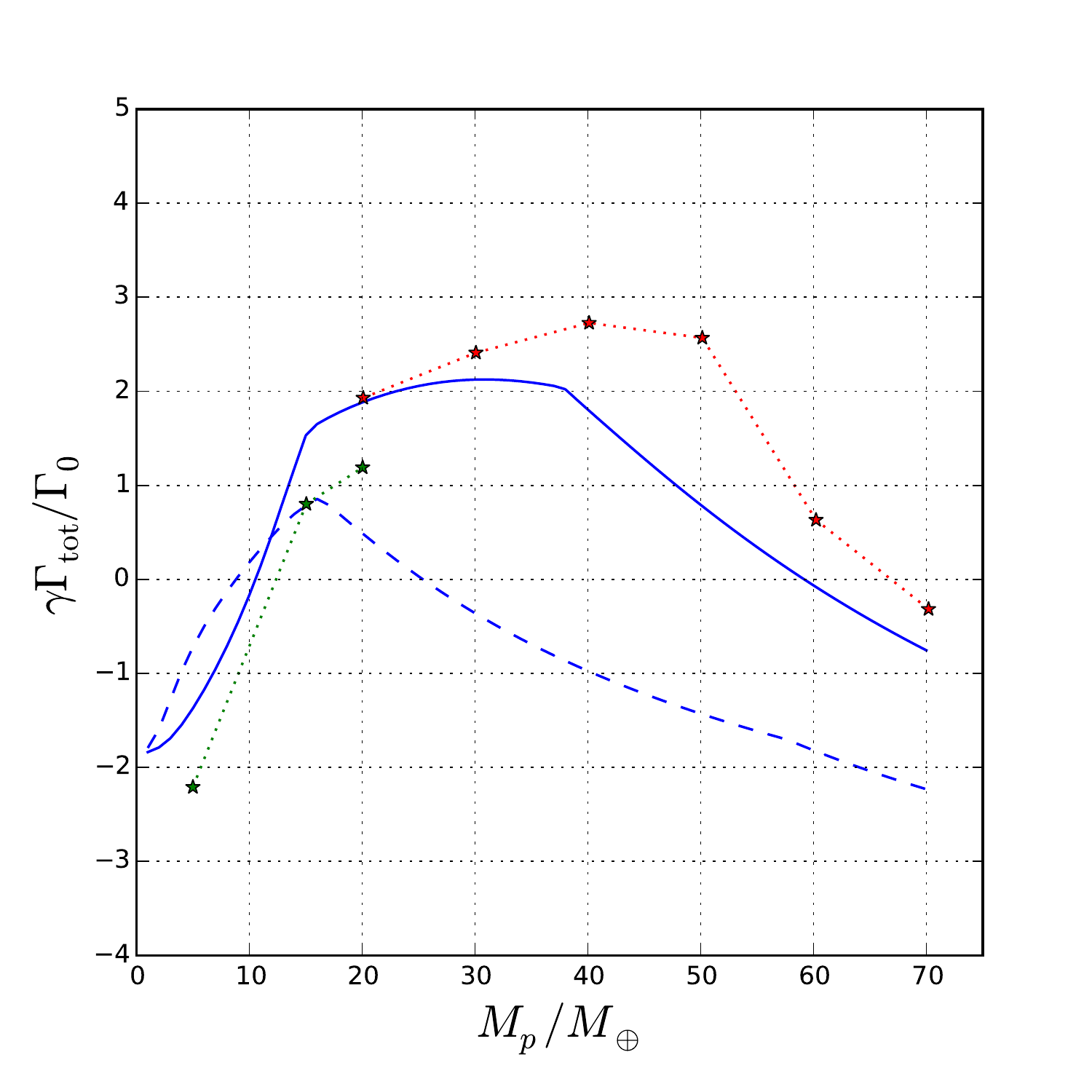}
  \includegraphics[width=.32\textwidth]{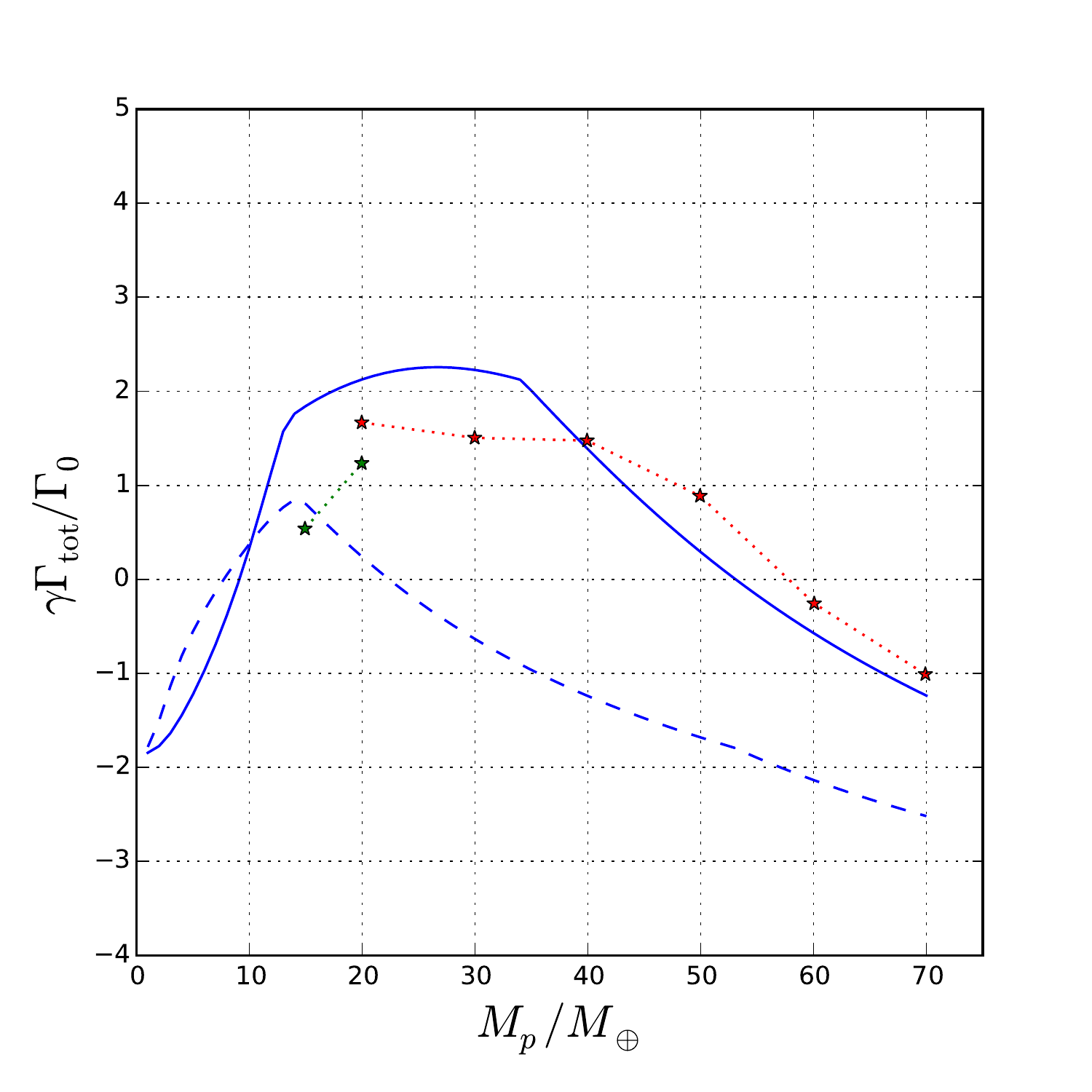}
  \includegraphics[width=.32\textwidth]{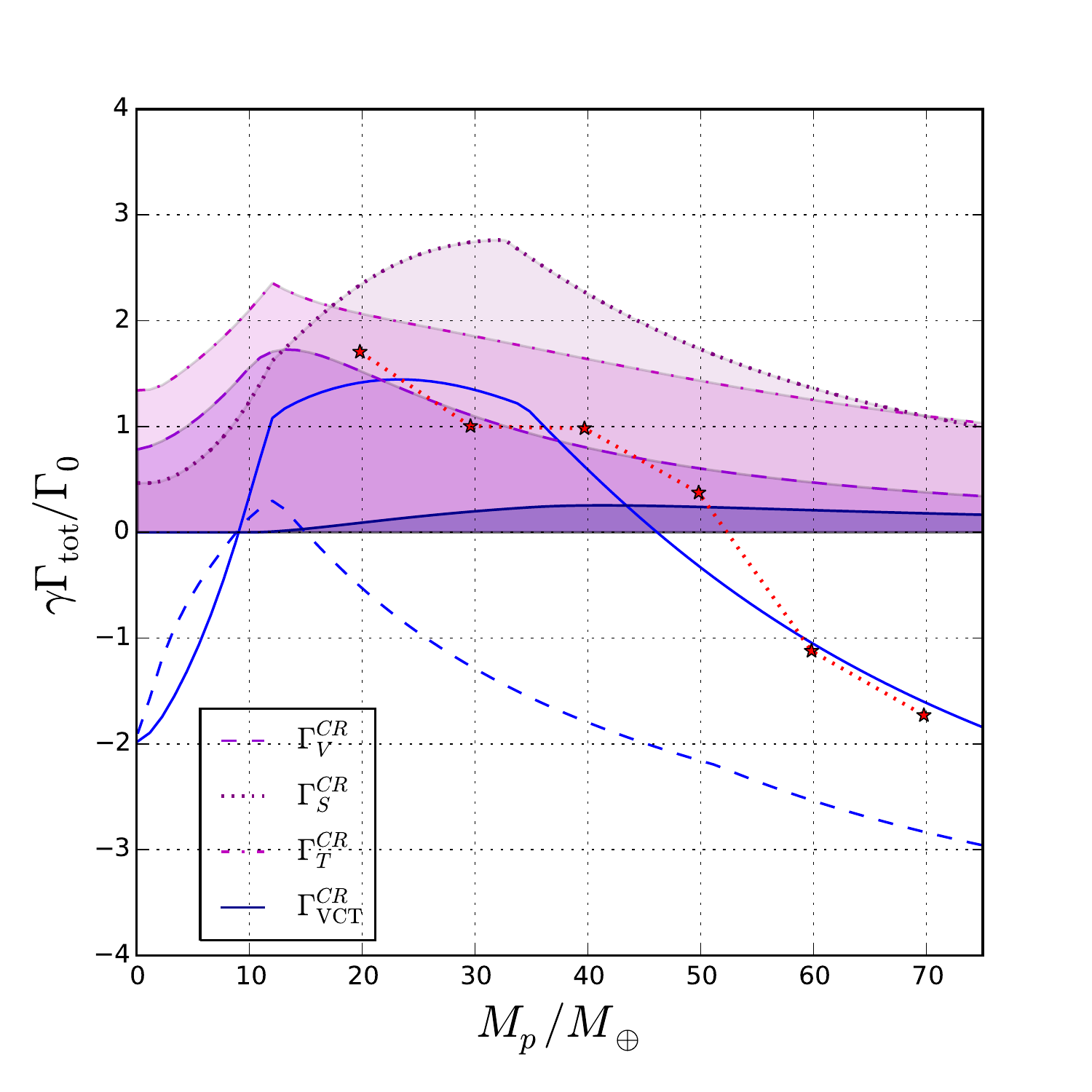}
  \caption{These figures show the comparison between our torque
    formula (solid line) and the results of
    \citet{2015MNRAS.452.1717L}. Plots from left to right should be
    compared to Figs.~5a to~5c of that work, and show results for a
    standard accretion disc (SAD) model. The dashed line corresponds
    to the outcome of our new torque formula when one uses the
    relationship $x_s=1.05\sqrt{q/h'}$ instead of
    Eq.~\eqref{eq:37}. This shows the importance of an accurate
    estimate of the width of the horseshoe region for intermediate
    mass planets (\emph{i.e.} here for planets above
    $10\;M_\oplus$). In the right plot we show the different
    components of the corotation torque, namely the vortensity torque
    (dashed line), the entropy torque (dotted line), the temperature
    torque (dash-dotted line) and the viscous coupling term (solid
    line). The latter is found to be virtually negligible, whereas the
    first three have similar orders of magnitude. The Lindblad torque
    is not represented, as it is constant and off the limits of the
    plot : $\gamma\Gamma_L/\Gamma_0\approx -4.6$. We thank Elena Lega
    for kindly providing us with the model data.}
  \label{fig:legacomp}
\end{figure*}

\begin{figure}
  \includegraphics[width=\columnwidth]{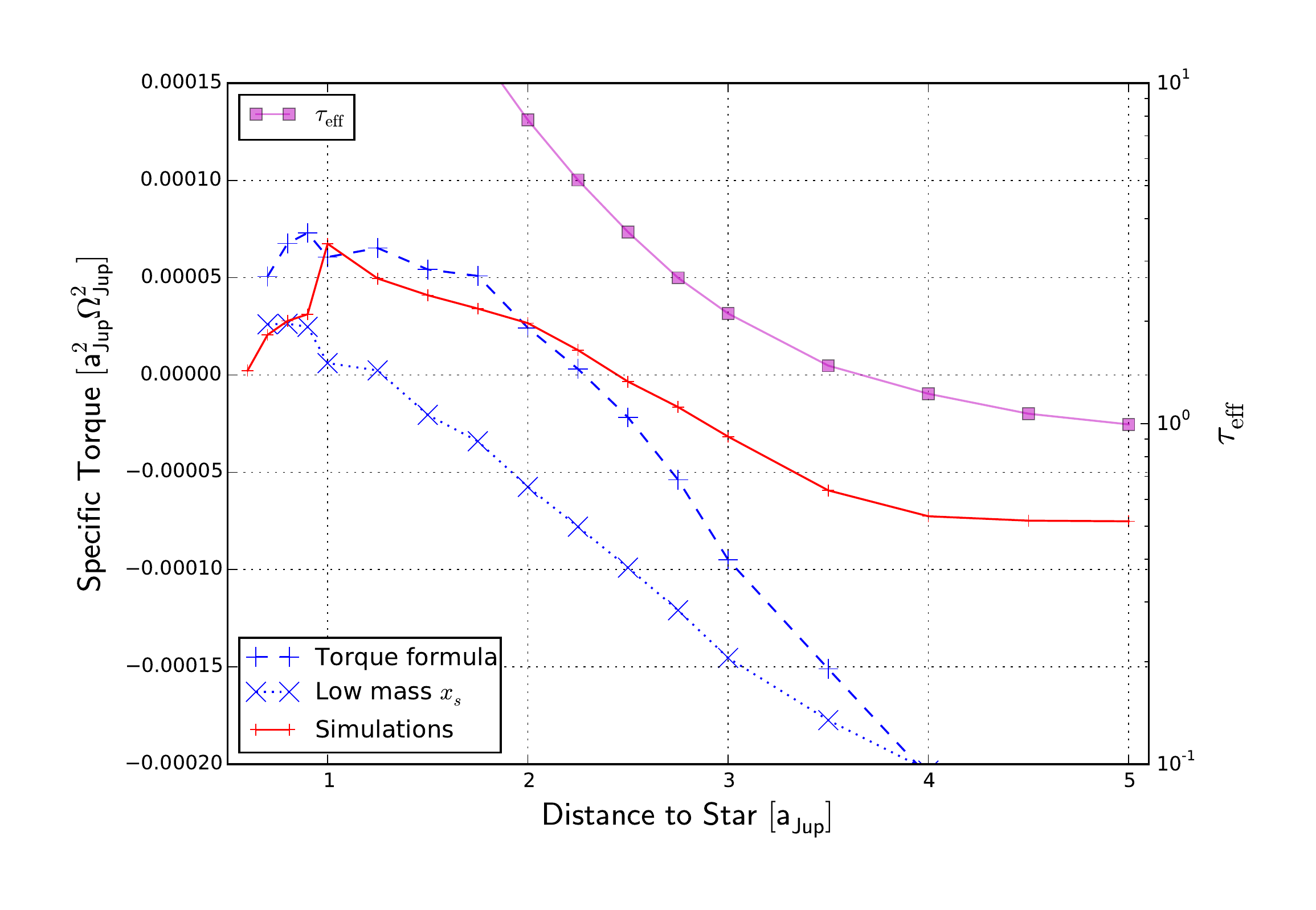}
  \caption{Comparison to the results of \citet[][hereafter
    BK11]{2011A&A...536A..77B} for a $20$~$M_\oplus$ planet. The
    dashed line shows the result of our torque formula, the solid line
    with crosses (in red in the electronic version) the results
    obtained by BK11, and the dotted line the outcome of our torque
    formula when one adopts the low-mass scaling for the horseshoe
    width. The right vertical axis shows the effective optical depth
    $\tau_\mathrm{eff}=\Sigma\kappa/2$ evaluated from the midplane
    quantities, and corresponds to the solid line with squares (in
    magenta in the electronic version). This figure should be compared
    to Figs.~2 and~3 of BK11. We thank Bertram Bitsch for kindly
    providing us with the model data.}
  \label{fig:bk}
\end{figure}

\section{Discussion}
\label{sec:discussion}
We compare here the torque value provided by our formula of
section~\ref{sec:new-torque-formula} to published studies of the
torque value in radiative discs, either as a function of the planetary
mass or of the orbital radius. Fig.~\ref{fig:legacomp} shows the
comparison to results recently published by
\citet{2015MNRAS.452.1717L}. They show a broad agreement between the
torque measured in their simulations and our torque formula. They also
show that the boost of the horseshoe width, described by
Eq.~\eqref{eq:37}, is an essential ingredient of the torque formula
above $10-15$~$M_\oplus$ (\emph{i.e.} for intermediate mass planets),
and that torque formulae based on a low-mass expression of the width
of the horseshoe region hardly yield a migration reversal.

It should be kept in mind that the torque measured in numerical
simulations is subjected to some inaccuracy, as can be seen from the
results for $20$~$M_\oplus$ (in the first two plots), for which two
torques values are displayed. This inaccuracy arises from a variety of
numerical effects, such as resolution (in particular the number of
zones over which the radial width of the horseshoe region is
resolved), or the recipe for the torque calculation (in particular
whether it includes all zones or whether it excludes those located
near the planet), etc.  Yet, some points or parts of the torque curves
are outliers with respect to our formula:
\begin{itemize}
\item At low planetary mass (on the left plot), the torque measured is
  somehow below the value predicted by our formula. This behaviour was
  already discussed by \citet{2015MNRAS.452.1717L}, and attributed to
  the ``cold-finger'' effect \citep{2014MNRAS.440..683L}, which is
  observed on low-mass planets when there is thermal diffusion in the
  disc. We note that \citet{2010ApJ...723.1393M} have obtained their
  Eq.~(156) ---~here Eq.~\eqref{eq:40}~--- using a fit of numerical
  simulations with different values of the thermal diffusivity, and
  may have unwittingly included a two-dimensional version of the
  ``cold-finger'' effect. It is therefore unclear to which extent the
  discrepancy found at low planetary mass is due to this effect. We
  also note that for the lowest planetary mass considered here
  ($5\;M_\oplus$), \citet{2014MNRAS.440..683L} found that the cold
  finger effect nearly vanishes. This issue requires further work, and
  will be presented elsewhere.
\item The plateau value for the mass range $15-35$~$M_\oplus$ on the
  middle plot, or the elbow of the left plot (at $38$~$M_\oplus$
  according to our formula, and at $\sim 50$~$M_\oplus$ in the results of
  \citet{2015MNRAS.452.1717L}) are also discrepant features, which are
  unlikely to be explained by the errors and systematic effects on the
  torque measurement. 
\end{itemize}
Beside the ``cold-finger'' effect, which might account for the mismatch
at low mass, there is a number of simplifications in our torque
formula which can explain the residual discrepancies between our
torque formula and the results of numerical simulations.  We neglect
the feed back of the planet's torque on the disc density profile. At
intermediate masses, this feed back tends to create a dip around the
orbit (precursor of the gap that would be carved at larger masses),
which has an effect on the different components of the torque. In
particular the effect can be different on the outer Lindblad torque
and on the inner Lindblad torque, resulting in a non-trivial effect on
the net Lindblad torque. Also, the slight decrease of density can
alter the thermal diffusivity at the planetary orbit, by virtue of
Eq.~\eqref{eq:34}. On the other hand, as we saw in
section~\ref{sec:diff-coeff-}, this equation is accurate only to
within $\sim 25$~\%, and both the saturation functions and the
blending coefficients between the linear torque and the horseshoe drag
depend sensitively on the value of the thermal diffusivity.  Also, we
use for these functions the analytic dependence worked out in two
dimensions by \citet{2010ApJ...723.1393M}, which can constitute
another reason for the discrepancy. Lastly, we have used for all
components of the corotation torque the actual adiabatic index of the
gas $\gamma$, instead of an effective adiabatic index as done for the
Lindblad torque \citep{2010ApJ...723.1393M,pbk11}. We comment that we
have performed runs with planets of mass $1$~$M_\oplus$ to
$20$~$M_\oplus$ embedded in radiative discs with different opacities
(namely $1.8$~cm$^2$.g$^{-1}$ and $5$~cm$^2$.g$^{-1}$). We have found
that the width of the horseshoe region at low mass corresponds indeed
to $x_s=1.05\sqrt{q/h}/\gamma^{1/4}$ rather than
$x_s=1.05\sqrt{q/h}/\gamma^{1/4}_\mathrm{eff}$. At larger masses, it
displays a growth above this value, as expected for the transition
from the low-mass to the high-mass branch of the horseshoe
width. Nonetheless, owing to thermal diffusion, the actual law might
be more complex than the one of Eq.~\eqref{eq:37} and this can have an
impact on the torque value, which is a very sensitive function of
$x_s$.

We finally compare the torque predicted by our formula to that
measured by \citet{2011A&A...536A..77B} in Fig.~\ref{fig:bk}. For
radii below $2.5$~$a_\mathrm{Jup}$, there is a satisfactory agreement
between the formula and the numerical simulations. Beyond that radius,
however, the predicted torque plummets to low values, whereas the
torque measured in the simulations plateaus at an intermediate,
negative value. We note that this happens when the disc's optical
depth drops below value of a few (at $r=a_\mathrm{Jup}$ it is
$\tau_\mathrm{eff}\sim 4$). Note that the actual optical depth of the
disc is somehow lower than the one displayed, as the opacity law
adopted by the authors was:
$\kappa/(1\;\mathrm{cm}^2\;\mathrm{g}^{-1})=2.0\cdot
10^{-4}[T/(1\;\mathrm{K})]^2$ (Bitsch, private communication). The
heat source of the disc being viscous friction, it is hotter at the
midplane, and therefore its opacity is largest at the midplane. This
example illustrates the fact that Eq.~\eqref{eq:34}, and therefore our
torque formula, are valid when the disc is optically
thick. Fig.~\ref{fig:bk} also shows that torque formulae based on the
low-mass scaling of the horseshoe width hardly predict any migration
reversal.

We comment that the dependence of the corotation torque on viscosity
and thermal diffusivity is entirely contained in the saturation
functions and blending coefficients. Since those are not reconsidered
in the present work, our torque has same dependence on viscosity and
thermal diffusivity as the dependence found by
\citet{2010ApJ...723.1393M}. Namely, Figs.~13, 17 and 18 of that work
show the dependence of the vortensity torque in the viscosity, and the
dependence of the entropy torque on the viscosity and thermal
diffusivity. The temperature term of the corotation torque was not
considered in that work, but it exhibits same dependence on viscosity
as the entropy torque.

\section{Conclusion}
\label{sec:conclusion}
We provide an updated formula for the torque experienced by a low mass
($q < 0.2h^3$) or intermediate mass planet
($0.2h^3\lesssim q\lesssim 2h^3$) in circular and non-inclined orbit
in an optically thick disc. Our torque formula agrees reasonably well
with the torque measured in numerical simulations, as shown by
comparison with recently published results. One key ingredient of the
formula is an accurate expression of the width of the horseshoe region
as a function of the planet-to-star mass ratio and of the disc's
aspect ratio.  Using in our torque formula the low-mass scaling
$x_s=1.05r_p\sqrt{q/h}$ clearly yields wrong, largely underestimated
results for intermediate mass planets. We have used tailored,
three-dimensional explorations of the parameter space to update
several numerical coefficients that appear in the torque formula. Our
work does not consider the effect found at low planetary mass by
\citet{2014MNRAS.440..683L} in discs with thermal diffusion, and
called by these authors the ``cold-finger'' effect. A further step
toward accurate torque formulae could be an analytic expression of
this torque component, as well as a relaxation of the different
simplifying assumptions mentioned in section~\ref{sec:discussion}.

\section*{Acknowledgements}
The authors thank Elena Lega and Bertram Bitsch for an extremely  
careful reading of a first draft of this manuscript, and for  
constructive feedback.  
Alejandra Jim\'enez acknowledges financial support from a UNAM DGAPA
fellowship. The simulations presented in this work were run on a GPU
cluster acquired with CONACyT's grant 178377. The authors acknowledge
support from UNAM's grant PAPIIT IN101616.

\bibliographystyle{mnras}

\bsp	
\label{lastpage}
\end{document}